\documentclass[
aps,%
12pt,%
final,%
notitlepage,%
oneside,%
onecolumn,%
nobibnotes,%
nofootinbib,%
superscriptaddress,%
noshowpacs
]%
{revtex4}

\usepackage{epsfig}
\usepackage{graphics}
\def\Ref#1{(\ref{#1})}
\begin{document}
	
	\title{SELF-CONSISTENT APPROACH TO THE DESCRIPTION OF\\
		RELAXATION PROCESSES IN CLASSICAL\\MULTIPARTICLE
		SYSTEMS}
	
	\author{\firstname{Anatolii~V.}~\surname{Mokshin}}

	\date{\today}
	
	\begin{abstract}
		The concept of time correlation functions is a very convenient theoretical tool in describing relaxation
		processes in multiparticle systems because, on one hand, correlation functions are directly related to experimentally measured quantities (for example, intensities in spectroscopic studies and kinetic coefficients
		via the Kubo-Green relation) and, on the other hand, the concept is also applicable beyond the equilibrium case. We show that the formalism of memory functions and the method of recurrence relations
		allow formulating a self-consistent approach for describing relaxation processes in classical multiparticle
		systems without needing a priori approximations of time correlation functions by model dependences and
		with the satisfaction of sum rules and other physical conditions guaranteed. We also demonstrate that
		the approach can be used to treat the simplest relaxation scenarios and to develop microscopic theories of
		transport phenomena in liquids, the propagation of density fluctuations in equilibrium simple liquids, and
		structure relaxation in supercooled liquids. This approach generalizes the mode-coupling approximation
		in the G\"{o}tze-Leutheusser realization and the Yulmetyev-Shurygin correlation approximations.

		
	\end{abstract}

	\maketitle
	
	\section{Introduction \label{Introduction}}

	Systems containing many interacting particles are the main subject of statistical mechanics, and describing such systems in both equilibrium and nonequilibrium is its main problem~\cite{Zubarev_UFN_1960}.  Among a variety of
	methods used to describe the \emph{dynamics} of multiparticle systems, the simple, useful, and fairly fruitful methods are based on using two-time advanced and retarded Green's functions introduced by Bogoliubov and
	Tjablikov~\cite{Plakida_1970,Tserkovnikov_1,Tserkovnikov_2,Tserkovnikov_3,Rudoi/Zubarev_UFN_1993,Bogolyubov/Tyablikov_1959,Vladimirov_2005,Cerkovnikov_TMF_2008}
and also on using the mathematical toolbox of time correlation functions (TCFs)
	~\cite{D_Zubarev_2002,Callen/Welton,R_Kubo_1957,Zwanzig_book_2001},
	distributions, moments, and cumulants~\cite{Klumov_UFN_2010,Lee3,Murtazaev_UFN_1999,I_Lifshitz_1982}.
 We do not discuss the full variety of existing methods
for describing the dynamics of multiparticle systems here (hardly feasible for obvious reasons, and not the
problem posed) and restrict ourself to considering the approach based on the concept of TCFs.
	
It is then very important that TCFs allow naturally realizing a \emph{statistical-probabilistic} treatment of
the system kinetics. Many experimentally measurable quantities are then expressed in terms of TCFs of the
corresponding variables. As examples, we mention kinetic coefficients (viscosity, diffusion, heat conduction)
and also responses measured in experiments in dielectric, optical, neutron, and X-ray spectroscopy and
in nuclear magnetic resonance. For example, Zwanzig~\cite{Zwanzig_book_2001} notes that the significance of TCFs in solving
problems in nonequilibrium statistical mechanics is comparable to the significance of partition functions and
spatial two-point correlation functions because many properties can be expressed directly in terms of TCFs~\cite{Zwanzig_book_2001}). In his statement, Zwanzig partly agrees with Berne and Harp~\cite{Berne_Adv_Chem_Phys}, who write in somewhat
greater detail that “time correlation functions have done for the theory of time-dependent processes what
partition functions have done for equilibrium theory. The time-dependent problem has become well defined,
but no easier to solve. One now knows which correlation function corresponds to a given time-dependent
phenomenon. Nevertheless, it is still extremely difficult to compute the correlation function''\footnote {Berne and Harp here refer primarily to analytic calculations of TCFs.} (p.67
	~\cite{Berne_Adv_Chem_Phys}). Although this was stated more than forty years ago, the relevance of this assertion is preserved for the simple
	reason of the lack of a  \emph{rigorous general} algorithm for calculating TCFs~\cite{Rudoi/Zubarev_UFN_1993}. It hence becomes necessary to
	apply various theoretical approximations to the relevant TCFs based on assumptions of a certain TCF
	behavior to be reproduced by model functions (e.g., exponential, Gaussian~\cite{Copley_Lovesey},
	hyperbolic secant~\cite{Tankeshwar1}, a combination thereof, and so on). Such approximations underlie Lovesey’s viscoelastic model~\cite{Copley_Lovesey},  the
	generalized hydrodynamic model~\cite{Hansen/McDonald_book_2006}, and the approximation of generalized collective modes~\cite{de_Schepp}. These
	approximations are applied to either the original TCF or some TCF that occurs on a certain level in an
	infinite chain of kinetic equations, which is essentially equivalent to terminating the chain.
	
Although the above approximation allows proposing some interpretation of the experimental results,
the absence of a manifest small parameter (e.g., as in the case of any dense liquids in the nonhydrodynamic
domain) makes it difficult to adduce significant physical arguments substantiating the applicability of such
approximations. It is then quite natural that interest is attracted to another, partly less rigorous approach
in which the behavior of the TCF and/or its spectral features are determined self-consistently~\cite{Olem_1995,Olem_1996}, without approximating the TCF by model functional dependences. This approach is based on theoretical
models developed in the framework of the method of recurrence relations~\cite{MHLee_2,MHLee_1,Lee3,Lee_Physica_Scr_1987,Lee_PRB_1982},
	the mode-coupling
	approximation~\cite{Gotze,Leutheusser_1984},  and the Yulmetev–Shurygin correlation approximations~\cite{Shurygin/Yulmetyev_TMF_1990,Yulmetyev3,Yulmetyev_P1,Yulmetyev_P2}.
Notably, this
approach forms a unified base for developing microscopic theories of different relaxation processes in multiparticle systems. Here, following Frenkel~\cite{J_Frenkel_book_1946},we restrict ourself to considering simple liquids, which we understand as one-component systems in which the particles\footnote{By the particles, we mean the structural elements of the system, i.e., atoms, molecules, colloid solution particles, etc. [35],whose dynamics satisfy the classical description [36]-[38].} are “identical” and interact via a spherical
potential.

	In Sec. 2, we present a \emph{general derivation}
of kinetic integro-differential equations in the framework of
projection operators. In Sec. 3, we discuss the method of recurrence relations, which, on one hand, can be
regarded as an alternative to the projection operator technique but, on the other hand, can serve as the
basis of a general self-consistent approach for describing relaxation processes in multiparticle systems. We
present this approach in Sec. 4, where we also demonstrate its application to the treatment of the simplest
relaxation scenarios (such as Gaussian relaxation and a damped oscillating correlator), to the development
of microscopic theories for calculating kinetic coefficients, and to the description of particle dynamics in
equilibrium and supercooled simple liquids.

	\section{Projection operators and dynamical correlations \label{Pr-operators}}
	
	We consider a system of $N$ identical classical particles of mass $m$, inside a volume $V$ with the number
	density $n=N/V$. The coordinates and velocities of the particles
	[$\vec{r}^{(l)}$ and $\vec{v}^{(l)}$, where $l=1,\ 2,\ \ldots, 3N$]
	constitute a $6N$-dimensional phase space
	\begin{equation} \label{eq: phase_space_Hilb}
	\mathcal{A}(\vec{r}^{(1)},\ \ldots, \vec{r}^{(l)},\; \;
	\vec{v}^{(1)},\ \ldots,\ \vec{v}^{(l)}),
	\end{equation}
on which dynamical variables are defined. For example, a dynamical variable can be chosen as the quantity
characterizing the particle trajectory
	\begin{equation} \label{eq: traj}
	\rho_s(\vec{r},t)=\delta(\vec{r}-\vec{r}_s(t)),
	\end{equation}
	the particle velocity $\vec{v}_s(t)$ or the particle energy
	
	\begin{equation}
	e_s(t) = \frac{m |\vec{v}_s|^2}{2} + \frac{1}{2}\sum_{s\neq j}^N
	U(r_{sj}),
	\end{equation}
	[$m$ -- is the particle mass and $U(r)$ -- is the interparticle interaction potential), the local density

	\begin{equation}
	\delta \rho(k)=\frac{1}{\sqrt{N}} \sum_{j=1}^{N}\textrm{e}^{i
		\vec{k} \cdot \vec{r}_{j}},
	\end{equation}
($\vec{k}$ is the wave vector), local order parameters;
	~\cite{Steinhardt}
	\begin{equation}
	q_{l}(s)= \left ( \frac{4\pi}{2l+1}  \sum_{m=-l}^{l}\left |
	\frac{1}{n_b(s)}\sum_{j=1}^{n_b(s)} Y_{lm}(\Omega_{sj}) \right |^2
	\right )^{1/2},\ \ \ l=6,\; 8,\; 10,\;\ldots ,
	\end{equation}
	[$n_b(s)$ --  is the number of particles within the first coordination sphere for particle $s$~\cite{Smirnov_UFN}]; or local shear
	stresses

	\begin{equation}
	\sigma_{xy}^{(\mathcal{N})} = \sum_{i=1}^{\mathcal{N}} \left ( m
	v_{ix}v_{iy} + \frac{1}{2} \sum_{i\neq j}^{\mathcal{N}} F_{ijx}
	r_{ijy} \right ),
	\end{equation}
	[$\mathcal{N}$ is the number of particles characterizing the local domain with
	 $\mathcal{N}\leq N$].

	In turn, the Hamiltonian $\hat{H}$ defines the averaging operation $\langle A\rangle$ in terms of the phase space distribution
	density $\rho \propto \exp [-(\hat{H}-\mu N)/(k_B
	T)]$,  where $\mu$ is the chemical potential. In the Hilbert space, we can then define a
	scalar product with the set of dynamical variables
	\begin{equation}
	(A, B)=\frac{\langle  A^*  B \rangle}{k_B T}, \ \langle  A \rangle
	= \langle  B \rangle =0. \label{scalar}
	\end{equation}
	In addition, the Hamiltonian defines the time evolution by the map $A \to A(t)$ via the canonical equation
	of motion
	\begin{eqnarray}\label{liuvil_eq}
	\frac{dA(t)}{dt}&=&i \{\hat{H},A(t)\} =i \hat{\mathcal{L}} A(t),\\
&&A\equiv A(t=0),  \nonumber
	\end{eqnarray}
	where $\{\hat{H},\ldots\}$ is the Poisson bracket and the Liouville operator
	\begin{equation}
	\hat{\mathcal{L}}=-i\sum_j
	\frac{\vec{p}_j}{m}\cdot\frac{\partial}{\partial \vec{r}_j} -i
	\sum_j \vec{F}_j \frac{\partial}{\partial \vec{p}_j}
	\end{equation}
	is Hermitian. A formal solution of Eq.~(\ref{liuvil_eq})
	has the form
	\begin{equation} A(t) =\textrm{e}^{i \hat{\mathcal{L}}t}A, \label{evol}
	\end{equation}
	where $\exp(i \hat{\mathcal{L}}t) $ is the evolution operator (or propagator) and $\vec{F}_j$ is the total force acting on the $j$particle.
	For the generality of the approach, we do not associate the variable $A$ with any concrete physical quantity
	at this stage in our presentation.
	
We define the TCF of two variables as:
	\begin{equation} \phi_{AB}(t,\tau)=\langle A^*(\tau) B(t+\tau) \rangle .
	\end{equation}
	The Hermiticity of the Liouville operator $\hat{\mathcal{L}}$ implies the identity
	\begin{equation} (\hat{\mathcal{L}}A, B ) \equiv (B,
	\hat{\mathcal{L}}A), \label{Herm}
	\end{equation}
	whence some properties of the TCF directly follow~\cite{Hansen/McDonald_book_2006}. First, using relation \Ref{Herm}, we can show that in the
	case of stationary processes, the equality
	\begin{eqnarray}
	\label{Stat} \langle A^*(\tau) B(t+\tau) \rangle &=& k_BT
	(\textrm{e}^{-i\hat{\mathcal{L}} \tau}A,
	\textrm{e}^{i\hat{\mathcal{L}} (t+\tau)}B)=\\ &=& k_BT (A,
	\textrm{e}^{i\hat{\mathcal{L}}t} B)= \langle A^*(0) B(t) \rangle .
	\nonumber
	\end{eqnarray}
	holds. Next, it follows from identity \Ref{Herm} that
	\begin{equation}
	\phi_{A\dot{B}}= k_BT ( A, i \hat{\mathcal{L}} B)  = - k_BT ( i
	\hat{\mathcal{L}} A, B)  = - \phi_{\dot{A}B}.
	\end{equation}
	
We restrict ourself to considering the stationary case, where we define the normalized time autocorrelation function
	\begin{equation}
	\phi_{AA}(t) = \frac{\langle
		A^*(0)A(t)\rangle}{\langle |A(0)|^2 \rangle} \equiv \frac{\langle
		A^*(0) \textrm{e}^{i  \hat{\mathcal{L}} t} A(0) \rangle}{\langle
		|A(0)|^2 \rangle}, \label{TCF_basic}
	\end{equation}
	with the properties:
	\begin{equation} \label{prop_TCF}  \phi_{AA}(t)|_{t=0}=1, \end{equation}
	\begin{equation} 1 \geq |\phi_{AA}(t)| \geq 0, \end{equation}
	\begin{equation} \label{eq: deriv}  \left . \frac{d \phi_{AA}(t)}{dt} \right |_{t=0} = 0. 
	\end{equation}
	Using the Wiener–Khinchin theorem~\cite{Rid}, we can show that the spectrum
	\begin{equation} \widetilde{\phi}_{AA}(\omega)=
	\int_{-\infty}^{+\infty}dt~\textrm{e}^{-i\omega t} \phi_{AA}(t)
	\end{equation} is real and positive,
	$\widetilde{\phi}(\omega)\geq 0$.
	
Expanding the time evolution operator in the right-hand side of  \Ref{evol} in a series, \begin{equation} A(t)= \left ( 1+ i
	\hat{\mathcal{L}} t -\frac{1}{2}\hat{\mathcal{L}}^2 t^2 + \ldots
	\right ) A(0),
	\end{equation} we obtain the short-time expansion of the TCF,
	\begin{equation} \phi_{AA}(t)= 1-
	\frac{1}{2!} \omega^{(2)}t^2 + \frac{1}{4!} \omega^{(4)} t^4 -
	\frac{1}{6!} \omega^{(6)} t^6 +\mathcal{O}(t^8), \label{taylor}
	\end{equation}
	where $\omega^{(p)}$ are the frequency moments of order $p$:
	\begin{eqnarray} \label{fr_m}
	\omega^{(2p)}&=&
	\left. (-i)^p\frac{d^{p}\widetilde{\phi}_{AA}(t)}{dt^p} \right |_{t=0},\\
	p&=&1,~ 2,~ \ldots ~ . \nonumber
	\end{eqnarray}

We define the projection operators
	\begin{equation} \Pi_{0} =
	\frac{A_0(0) \rangle\langle A_0^*(0)}{\langle |A_0(0)|^2\rangle},
	\ \ \ \ P_0=1-\Pi_0, \label{R_a}
	\end{equation}
	which have the properties
	\begin{equation} \Pi_0 + P_0=1,
	\end{equation}
	\begin{equation}
	\Pi_0^2=\Pi_0,\ \ \
	P_0^2=(1-\Pi_0)^2=P_0,
	\end{equation}
	\begin{equation}
	\Pi_0 A_0(0)=A_0(0), \label{qq}
	\end{equation}
	\begin{equation}
	\Pi_0 A_0(t)= A_0(0)F(t),
	\end{equation}
	\begin{equation}
	\Pi_0 P_0=\Pi_0(1-\Pi_0)= P_0 \Pi_0 =0.
	\end{equation}
	
	We act with the operators $\Pi_0$ and $P_0$ on Liouville equation\Ref{liuvil_eq}
from the left. For convenience, we
introduce the notation
	\begin{equation} A_0^{'}(t)=\Pi_0 A_0(t),\ A_0^{''}(t)=P_0 A_0(t).
	\label{shtrih}
	\end{equation}
Liouville equation~(\ref{evol}) then becomes:
	\begin{equation} \left . 
	\begin{array}{l}
	\displaystyle{\frac{d}{dt}} \Pi_0 A_0(t)=
	\displaystyle{\frac{d}{dt}} A_0^{'}(t)=i
	\hat{\mathcal{L}}_{11}^{0}A_{0}^{'}(t) + i
	\hat{\mathcal{L}}_{12}^{0}A_{0}^{''}(t), 
	\vspace{0.5cm} \\
	\displaystyle{\frac{d}{dt}} P_0 A_0(t)=
	\displaystyle{\frac{d}{dt}} A_0^{''}(t)=i
	\hat{\mathcal{L}}_{21}^{0}A_{0}^{'}(t) + i
	\hat{\mathcal{L}}_{22}^{0}A_{0}^{''}(t),
	\end{array}
	\right.  \label{system}
	\end{equation}
where we set
	\begin{equation}
	\hat{\mathcal{L}}_{11}^{0}=\Pi_0 \hat{\mathcal{L}} \Pi_0,\
	\hat{\mathcal{L}}_{12}^{0} = \Pi_{0} \hat{\mathcal{L}} P_{0},\
	\hat{\mathcal{L}}_{21}^{0} = P_{0} \hat{\mathcal{L}} \Pi_{0},\
	\hat{\mathcal{L}}_{22}^{0} = P_{0} \hat{\mathcal{L}} P_{0}.
	\label{reduc}
	\end{equation}
	\begin{equation}
	\hat{\mathcal{L}}_{11}^{0}+ \hat{\mathcal{L}}_{12}^{0}+
	\hat{\mathcal{L}}_{21}^{0}+
	\hat{\mathcal{L}}_{22}^{0}=\hat{\mathcal{L}}.
	\end{equation}
	
	Applying the Laplace transformation operator $\hat{LT}$ to the second equation of system
	\begin{equation} \label{eq: LT}
	\hat{LT}[f(\tau)] = \int_{0}^{\infty} e^{-s\tau}f(\tau)d\tau =
	\tilde{f}(s)
	\end{equation}
	we obtain
	\begin{equation} \widetilde{A}_0^{''}(s)=
	\frac{A_0^{''}(0)}{s- i\hat{\mathcal{L}}_{22}^0} +
	\frac{i\hat{\mathcal{L}}_{21}^0 \widetilde{A}_0^{'}(s)}{s- i
		\hat{\mathcal{L}}_{22}^{0}} . \label{immed}
	\end{equation}
Doing the inverse Laplace transformation in Eq \Ref{immed},
we obtain
	\begin{equation}
	A_{0}^{''}(t) =\textrm{e}^{i
		\hat{\mathcal{L}}_{22}^{0}t}A_{0}^{''}(0) +i \int_{0}^{t} d\tau
	\textrm{e}^{i  \hat{\mathcal{L}}_{22}^{0}\tau}
	\hat{\mathcal{L}}_{21}^{0} A^{'}(t - \tau). \label{solution1}
	\end{equation}
	
	In accordance with property \Ref{qq} we have
	\begin{equation} A_{0}^{''}(0)=
	P_{0}A_0(0)=(1-\Pi_0)A_{0}(0) =0. \label{ravn}
	\end{equation}
	Substituting Eq. \Ref{solution1} in the first equation in system
	\Ref{system}, and using \Ref{ravn}  we obtain the integro-differential
	equation:
	\begin{equation}
	\begin{array}{l}
	\displaystyle{\frac{d}{dt}} A_0^{'}(t)= i
	\hat{\mathcal{L}}_{11}^{0}A_{0}^{'}(t) -
	\hat{\mathcal{L}}_{12}^{0}
	\int_{0}^{t} d\tau \textrm{e}^{i  \hat{\mathcal{L}}_{22}^{0}\tau} \hat{\mathcal{L}}_{21}^{0} A_{0}^{'}(t-\tau)= \\
	\ \ \ \ = i  \hat{\mathcal{L}}_{11}^{0}A_{0}^{'}(t) - \int_{0}^{t}
	d\tau \hat{\mathcal{L}}_{12}^{0} \textrm{e}^{i
		\hat{\mathcal{L}}_{22}^{0}\tau} \hat{\mathcal{L}}_{21}^{0}
	A_{0}^{'}(t-\tau).
	\end{array}  \label{solution2}
	\end{equation}
	Representing the operator $\Pi_{0}$  in the form
	\begin{equation}
	\Pi_{0}=R_{0}S_{0},\ R_{0}=A_{0}(0) \rangle,\ S_{0}=\frac{\langle
		A_{0}^{*}(0)}{\langle |A_{0}(0)|^{2} \rangle}, \label{pi_a}
	\end{equation}
we find the properties of the operator $S_0$:
	\begin{equation}
	S_{0}A_{0}(0)=1,\ S_{0}A_{0}(t)=\phi_{AA}(t),\ S_{0}R_{0}=1,\
	S_{0}\Pi_{0}=S_0. \label{propert}
	\end{equation}
	We act with $S_0$ from the left consecutively on each of the three parts of expression \Ref{solution2}. Then the left-hand
	side of
	\Ref{solution2} becomes:
	\[
	S_0
	\frac{d}{dt}A_0^{'}(t) =\frac{d\phi_{AA}(t)}{dt}, \label{first_1}
	\]
	and the first term in the right-hand side of \Ref{solution2} is
	\[
	i S_0 \hat{\mathcal{L}}_{11}^{0} A_{0}^{'}(t)= i S_{0}
	\hat{\mathcal{L}} R_{0} \phi_{AA}(t). \label{first_2}
	\]
	The integrand in  \Ref{solution2} is
	\[ S_0 \hat{\mathcal{L}}_{12}^{0}\textrm{e}^{i
		\hat{\mathcal{L}}_{22}^{0}\tau} \hat{\mathcal{L}}_{21}^{0}
	A_{0}^{'}(t-\tau)
	= \frac{\langle (A_0(0) \hat{\mathcal{L}})^{*} \textrm{e}^{i
			\hat{\mathcal{L}}_{22}^{0}\tau}(\hat{\mathcal{L}} A_{0}(0))
		\rangle}{\langle |A_0(0)|^2\rangle} - \left [ \frac{\langle
		A_0^{*}(0) \hat{\mathcal{L}} A_{0}(0) \rangle}{\langle
		|A_0(0)|^2\rangle} \right ]. \label{first_3}
	\]
	We find the equation of motion
	\begin{eqnarray} \label{eq: GLE_ZM}
	\frac{dA_0(t)}{dt} = - \int_0^{\infty} A_0(t-\tau) \frac{\langle
		(i\hat{\mathcal{L}}A(0))^*A_1(\tau) \rangle}{\langle|A_0(0)|^2
		\rangle}d\tau +A_1(t),
	\end{eqnarray}
which for the TCF becomes the kinetic integro-differential equation
	\begin{eqnarray} \label{first_eq_1}
	\frac{d\phi_{0}(t)}{dt}= 
	&-&\Delta_{1}\int_{0}^{t} \phi_1(\tau) \phi_{0}(t-\tau) d\tau ,\\
	\phi_{0}(t) & \equiv & \phi_{AA}(t), \nonumber
	\end{eqnarray}
	The quantity
	\begin{equation}
	A_{1}(t)= i \hat{\mathcal{L}} A_{0}(t),\label{second_var}
	\end{equation}
is the second dynamical variable orthogonal to $A_0$:
	\[
	(A_0,A_1)=0.
	\]
	The relation
	\begin{equation}
	\Delta_{1}=\frac{\langle |A_1(0)|^2\rangle}{\langle
		|A_0(0)|^2\rangle} \label{Om_1_1}
	\end{equation}
defines the so-called first frequency parameter, which has the dimension of frequency squared. As can be
seen from~\Ref{second_var} and
	\Ref{Om_1_1},  the explicit form of the frequency parameter $\Delta_1$ depends on the quantity chosen
	as the original variable $A_0$.
	Finally,
	\begin{equation}
	\phi_1(\tau)=\frac{\langle A_1^*(0)\textrm{e}^{i
			\hat{\mathcal{L}}_{22}^{0}\tau}A_1(0) \rangle}{\langle
		|A_1(0)|^2\rangle} \label{first_memory}
	\end{equation}
	is the so-called first-order memory function, which is a TCF of the variable $A_1$ and has all the properties
	in  \Ref{prop_TCF}.
	
	The sequence of procedures \Ref{shtrih}--\Ref{first_eq_1}
is a standard derivation of a kinetic integro-differential equation
for the TCF $\phi_0(t)$ from the equation of motion for a dynamical variable $A_0$ using projection operators. It
is essential here that Eq.\Ref{first_eq_1}was derived exactly. Such a derivation was first done by Zwanzig~\cite{Zwanzig} and
Mori~\cite{Mori}, and a similar realization of the method of projection operators later came to be known as the
formalism of memory functions (or the Zwanzig–Mori formalism)~\cite{Hansen/McDonald_book_2006}. We note that $A_1$ is also a dynamical
variable defined on phase space~\Ref{eq: phase_space_Hilb}, and its evolution is given by the corresponding Liouville equation of
form
	\Ref{liuvil_eq}.  Defining projection operators for $A_1$
	similarly to \Ref{R_a}  and repeating the sequence of actions
	in
	\Ref{shtrih}--\Ref{first_eq_1}, we obtain a kinetic integro-differential equation for the TCF $M_1(t)$, which in turn involves a
	second-order memory function
	$\phi_2(t)$, which is the TCF of a new variable $A_2$, orthogonal to
	$A_0$ and $A_1$.

	The technique of projection operators allows separating the
	"proper motion" of the chosen dynamical
	variable
	$A_0$ and the effects due to the interaction of $A_0$ with a "flux" variable orthogonal to it, whose
	dynamics is related to other degrees of freedom. For example, in the case where $A_0$
	is the particle velocity $A_1$
	is related to a “stochastic force” that affects the motion of the particle via its interaction with the
	environment. The quantity
	$A_1$ directly enters the equation of motion for the original dynamical variable $A_0$.  Another contribution that also determines the evolution of $A_0$, takes the correlations in the dynamics of
	$A_1$ and its interaction with the TCF $\phi_0(t)$ into account.  In accordance with this formulation, the behavior
	of $A_0(t)$ can become more regular as a result of the interaction between $A_0$ and $A_1$, that has occurred at the
	preceding instants. The mechanism of this contribution, governing the system behavior and contributing to dissipation, is associated with the so-called statistical memory ~\cite{Zwanzig_book_2001},\cite{Mokshin/Yulmetyev},\cite{Mokshin/Yulmetyev_book_2006}.
	
As a result, the process is described in the orthogonal basis
	\begin{equation} \label{eq: ort_basis}
	\mathbf{A}=\{A_0,\; A_1,\; A_2,\; \ldots,\; A_{\nu},\; \ldots\},
	\end{equation}
	where the variables have the property
	\begin{equation}
	\langle A_{\nu} A_{\mu} \rangle = \left\{
	\begin{array}{l}
	0,\ \hspace{1.0cm} \nu \neq \mu, \\
	\langle | A_{\nu}|^2 \rangle,\  \nu = \mu,
	\end{array}
	\right.  \label{ortog}
	\end{equation}
	\[
	\\ \nu,\; \mu = 0,\;1,\; 2,\; \ldots
	\]
and the basis construction procedure is identical to the Gram–Schmidt orthogonalization procedure~\cite{Rid}. The variables in the set
	$\mathbf{A}$ are related by the chain of integro-differential equations
	\begin{equation} \label{eq: chain_variable}
	\frac{dA_{\nu}(t)}{dt} = - \Delta_{\nu+1} \int_0^{\infty}
	A_{\nu}(t-\tau) \frac{\langle A_{\nu+1}(0))^*A_{\nu+1}(\tau)
		\rangle}{\langle|A_{\nu+1}(0)|^2 \rangle}d\tau +A_{\nu+1}(t),
	\end{equation}
	\begin{equation}
	\Delta_{\nu+1}=\frac{\langle |A_{\nu+1}(0)|^2\rangle}{\langle
		|A_{\nu}(0)|^2\rangle} \label{eq: freq_par},
	\end{equation}
which corresponds to the chain of kinetic integro-differential equations
	\begin{equation} \label{eq: chain_TCF}
	\frac{d\phi_{\nu}(t)}{dt} = - \Delta_{\nu+1} \int_0^{\infty}
	\phi_{\nu}(t-\tau) \phi_{\nu+1}(\tau) d\tau .
	\end{equation}
	
	The original process associated with the variable
	$A_0$, 0 is no longer singled out here and is characterized
	by its interrelation with other processes occurring in the system. This can be seen especially clearly in the
	method of recurrence relations~\cite{MHLee_1},\cite{MHLee_2}, proposed by Lee~\cite{Lee_PRB_1982}.

	\section{Method of recurrence relations \label{RRM}}
	
An important feature of the method of recurrence relations is that the system dynamics are here directly
related to the properties of the relevant Hilbert space containing specific physical information~\cite{Mokshin/Yulmetyev_book_2006,Lee_PRB_1982}.  For
example, the behavior of a selected dynamical variable is directly determined by the dimension and shape
of the spatial and structure characteristics. Similar ideas were expressed in works of Bogoliubov\cite{Bogolyubov_book,Bogolyubov_1973},
	Zubarev~\cite{D_Zubarev_2002}, and Kubo ~\cite{R_Kubo_1957}.
	
	\subsection{Basic recurrence relations.}
	
	Let the space $\mathcal{S}$ be the realization of an abstract Hilbert space  \cite{MHLee_2,MHLee_1}, defined by its scalar product, also known as the Kubo product~
	\cite{R_Kubo_1957}
	\begin{equation} \label{scal}
	(X,Y)= k_B T \int_0^{\beta} \langle \exp(\lambda \hat{H})Y^\dag
	\exp(-\lambda \hat{H})X \rangle d\lambda,
	\end{equation}
	where $X,~Y \subset \mathcal{S}$ and $(X,Y)$ is called the relaxation function of $X$ and $Y$. Here $\langle X Y
	\rangle$ denotes averaging
	over the ensemble, defined by the formula
	\begin{equation}
	\langle X Y \rangle = \frac{\mathrm{Tr}[XY \exp(- \hat{H}/k_B
		T)]}{\mathrm{Tr}[\exp(- \hat{H}/k_B T)]}.
	\end{equation}

The dynamical variable $A_0(t)$  is a vector in this space and has the norm $||A_0(t)||=
	(A_0(t),A_0(t))$. If
	the system is Hermitian, then the norm does not change during motion over the surface of the space $\mathcal{S}$, i.e.,
	$||A_0(t)||= ||A_0||$. Hence, $A_0(t)$  can change only direction, but the length $||A_0(t)||$ must remain the same. As
	$t$ increases, the vector $A_0(t)$ describes a trajectory on the surface of
	$\mathcal{S}$ defined by the properties of this
	space~\cite{MHLee_1}.
	
	The volume of $\mathcal{S}$ is determined by a full set of orthogonal basis vectors $\{f_0$, $f_1$, $\ldots$,
	$f_{d-1}\}$,
	\begin{eqnarray}
	(f_{\nu}(k),f_{\mu}(k))&=&(f_{\nu}(k),f_{\nu}(k))
	\delta_{\nu\mu},\\  \nu,\ \mu &=&0,~1,~\ldots,~ d-1, \nonumber
	\end{eqnarray}
	which are linked by the so-called  \emph{first recurrence relation}:
	\begin{eqnarray}
	\label{RR1} f_{\nu+1}&=&i\mathcal{L}f_{\nu}
	+\Delta_{\nu}f_{\nu-1},  \\
	\Delta_{\nu}&=&\frac{(f_{\nu},f_{\nu})}{(f_{\nu-1},f_{\nu-1})},\nonumber\\
	\nu&=&0,~1,~2,~3,~\ldots,~d-1,\nonumber \\
	\ \ f_{-1}&=&0,\ \ \ \ \Delta_{0}\equiv1. \nonumber
	\end{eqnarray}
	The quantity $d$ defines the dimension of the space. The basis vectors depend on the general properties of
	the scalar product for that space and are therefore physical characteristics of the system.
	
	At an arbitrary instant $t$ the vector $A_0(t)$  has projections on each basis vector $f_{\nu}$. We let $\phi^{(\nu)}(t)$
	denote the magnitude of the projection of $A_0(t)$ on the $\nu$-th basis vector at an instant $t$
	:
	\begin{equation} \label{eq: RRA_TCF}
	\phi^{(\nu)}(t)=\frac{(A_0(t),f_{\nu})}{(f_{\nu},f_{\nu})}.
	\label{rel_fun}
	\end{equation}
	Following~\cite{Lee_PRB_1982}, we can write an orthogonal decomposition for $A_{0}(t)$:
	\begin{equation}
	A_0(t)=\sum_{\nu=0}^{d-1}\phi^{(\nu)}(t)f_{\nu} \label{RRA1}
	\end{equation}
	and define the initial condition for the dynamical variable
	\begin{equation} f_0=A_0(t=0)=A_0. \label{f_0}
	\end{equation}
	From Eq. \Ref{RRA1} we then obtain the boundary conditions for $\phi^{(\nu)}(t=0)$:
	\begin{equation}
	\phi^{(\nu)}(t=0)=\left\{
	\begin{array}{rcl}
	1, && \nu=0,\\
	0, && \nu= 1,~2,~3,~\ldots.\\
	\end{array}
	\right.
	\end{equation}
	
	Comparing expressions (\ref{scalar}), (\ref{TCF_basic}) and
	\Ref{rel_fun} we find that $\phi^{(0)}(t)$ can be identified with the relaxation function
	\begin{equation}
	\phi^{(0)}(t) \equiv \phi_0(t),
	\end{equation}
	where
	\begin{eqnarray} \label{eq: rel_function}
	\phi_0(t)=\frac{(A_0(t),f_{0})}{(f_{0},f_{0})}=
	\frac{(A_0(t),A_0(0))}{(A_0(0),A_0(0))}.
	\end{eqnarray}
	
	From \Ref{RR1} and (\ref{eq: rel_function}) we can now obtain the so-called \emph{second recurrence relation }
	\cite{MHLee_1,MHLee_2}:
	\begin{eqnarray} \label{RR2} \Delta_{\nu+1}
	\phi^{(\nu+1)}(t) &=&-\frac{d \phi^{(\nu)}(t)}{dt}
	+\phi^{(\nu-1)}(t),
	\\
	\nu&=&0,\ 1,\ 2,\ \ldots \ d-1, \nonumber \\
	& & \phi^{(-1)}\equiv 0. \nonumber
	\end{eqnarray}
	Hence, while first recurrence relation ~\Ref{RR1}
establishes a relation between dynamical variables, second relation~\Ref{RR2}  signifies a connection between the dynamics of different relaxation processes occurring in the
system.

	\subsection{Continued fractions and integro-differential equations.}
	
	Applying the Laplace transformation operator  $\hat{LT}$ to second recurrence relation~(\ref{RR2}), we can obtain two equations
	\begin{eqnarray} \label{rec_w}
	1&=& s \widetilde{\phi}^{(0)}(s)+\Delta_1 \widetilde{\phi}^{(1)}(s), \\
	\widetilde{\phi}^{(\nu-1)}(s)&=& s \widetilde{\phi}^{(\nu)}(s)+
	\Delta_{\nu+1} \widetilde{\phi}^{(\nu+1)}(s),\nonumber \\
	\nu&=&0,\ 1,\ 2,\ \ldots \ d-1,\nonumber
	\end{eqnarray}
	the simultaneous solution of which gives the continued fraction
	\begin{equation}
	\widetilde{\phi}^{(0)}(s) \equiv \widetilde{\phi}_{0}(s)=
	\frac{1}{\displaystyle s+\frac{\Delta_{1}}{\displaystyle
			s+\frac{\Delta_{2}}{\displaystyle
				s+\frac{\Delta_{3}}{\displaystyle s+\ddots}}}}.
	\label{inf_fraction1}
	\end{equation}
	
	Following~\cite{MHLee_1}, we can define the quantity
	\begin{equation}
	\widetilde{\psi}^{(\nu)}(s)=\frac{\widetilde{\phi}^{(\nu)}(s)}{\widetilde{\phi}^{(0)}(s)},\
	\ \  \nu=1,\ 2,\ \ldots \ d-1. \label{b1}
	\end{equation}
	Using relations\Ref{rec_w} and taking \Ref{b1} into account, we obtain
	\begin{equation}
	\widetilde{\phi}^{(0)}(s)= \frac{1}{s+\Delta_1
		\widetilde{\psi}^{(1)}(s)}. \label{b1_w}
	\end{equation}
	From \Ref{b1_w} and  \Ref{inf_fraction1} we then find
	\begin{equation}
	\widetilde{\psi}^{(1)}(s) \equiv \widetilde{\phi}_{1}(s) =
	\frac{1}{\displaystyle s+\frac{\Delta_{2}}{\displaystyle
			s+\frac{\Delta_{3}}{\displaystyle
				s+\frac{\Delta_{4}}{\displaystyle s+\ddots}}}}.
	\label{inf_fraction2}
	\end{equation}
	Applying the inverse Laplace transformation to \Ref{b1_w},
	yields
	\begin{equation}
	\frac{d \phi^{(0)}(t)}{dt}+\Delta_1 \int_{0}^{t}d\tau ~
	\psi^{(1)}(t-\tau)\phi^{(0)}(\tau)=0. \label{non_M_R}
	\end{equation}
	Equation \Ref{non_M_R} corresponds to the first integro-differential equation in chain~(\ref{eq: chain_TCF})
	[also see Eq.~(\ref{first_eq_1})], where $\psi^{(1)}(t)$
	is associated with the first-order memory function $\phi_1(t)$,
	\[
	\phi_1(t) \equiv \psi^{(1)}(t).
	\]
	
	The function $\psi^{(1)}$ is a projection of the variable
	$A_1(t)$ on the basis vector $f_1$,
	\begin{eqnarray}
	\psi^{(1)}(t)=\frac{(A_1(t),f_{1})}{(f_{1},f_{1})} =
	\frac{(A_1(t),A_1(0))}{(A_1(0),A_1(0))}, \label{rel_funB}
	\end{eqnarray}
	where
	\begin{equation}
	f_1=A_1(t=0)=A_1.
	\end{equation}
	Similarly to decomposition  \Ref{RRA1} we can write a decomposition for $A_1(t)$ \cite{MHLee_1,MHLee_2}
	\begin{equation}
	A_1(t)=\sum_{\nu=1}^{d-1}\psi^{(\nu)}(t)f_{\nu}, \label{RRA1B}
	\end{equation}
	where the function $\psi^{(\nu)}(t)$ determines the magnitude of the projection of the vector
	$A_1$ on the basis $f_{\nu}$ at
	an instant $t$,
	\begin{eqnarray}
	\psi^{(\nu)}(t)&=&\frac{(A_1(t),f_{\nu})}{(f_{\nu},f_{\nu})}, \ \
	\ \nu=1,~2,~3,~\ldots , \label{rel_funBB}
	\end{eqnarray} and also satisfies the boundary conditions
	\begin{equation}
	\psi^{(\nu)}(t=0)=\left\{
	\begin{array}{rcl}
	1, && \nu=1,\\
	0, && \nu= 2,~3,~4,~ \ldots.\\
	\end{array}
	\right.
	\end{equation}
	Hence, if the dynamics of $A_0$ occur in the space $\mathcal{S}$, spanned by basis vectors
	$f_0,~f_1,~f_2,~\ldots,~f_{d-1}$, then the
	dynamics of $A_1$ --
	occur in the subspace $\mathcal{S}_1$, spanned by the basis vectors
	$f_1,~f_2,~f_3,~\ldots,~f_{d-1}$ and $\mathcal{S}_1\perp f_0$.
	
	The representation of $\widetilde{\phi}_0(s)$ as a continued fraction, Eq.~(\ref{inf_fraction1}),  derived in the method of recurrence
	relations, is identical to the representation in the frequency Laplace map of the chain of kinetic integrodifferential equations~(\ref{eq: chain_TCF}), obtained in the Zwanzig–Mori formalism (see Sec.~\ref{Pr-operators}).

	\section{Self-consistent approach\label{sec: selfconsist}}
	
The Zwanzig–Mori formalism suggests a possibility of describing the relaxation processes in multiparticle systems using a system of integro-differential equations. But the formalism does not offer any
general algorithm for finding solutions of these equations. Hence, various theoretical approximations keep
appearing based on presuming some particular behavior of the TCF to be reproduced by model functions:
exponential~\cite{Mori}, Gaussian ~\cite{Copley_Lovesey},
	hyperbolic secant ~\cite{Tankeshwar1}, combinations thereof, and other model functions.
	Such approximations underlie the viscoelastic Lovesey model  (Lovesey)~\cite{Copley_Lovesey}, the so-called generalized hydrodynamic
	model~\cite{Balucani_Zoppi}, and the approach of generalized collective modes~\cite{de_Schepp}.An entirely natural question to be asked
	here is whether a general self-consistent approach for finding the TCF $\phi_{\nu}(t)$ can be developed without
	any recourse to guesswork and approximation of the relaxational behavior by some model dependences.

The need for developing such an approach is obvious because a priori approximations of the relaxational
behavior of the TCF $\phi_{\nu}(t)$ frequently conflict with a number of physical principles. For example
	\begin{enumerate}
		\item the TCF spectral characteristics must satisfy sum rules~\cite{Temperly2};  \item the TCF must take finite values as follows from \Ref{prop_TCF}-\Ref{eq: deriv}, and; \item a rigorous and well-defined description of the dynamics is possible in an orthogonal basis~\cite{Shurygin/Yulmetyev_TMF_1990},\cite{Rid}.
	\end{enumerate}
	
	On the other hand, the method of recurrence relations
	(see Sec.\ref{RRM}) suggests that the dynamics of the
	investigated system are completely determined by the frequency parameters  $\Delta_{\nu}$ (in particular, the structure
	characteristics determining the values of the frequency parameters).  Hence, the problem of finding the
	original TCF $\phi_0(t)$  is reducible to analytically calculating the frequency parameters $\Delta_{\nu}$, and estimating
	their values.
	 Because the frequency parameters are directly related to dynamical variables [see~\Ref{eq: freq_par}], their
	 number and values are determined by physical conditions in which the relaxation process associated with
	 the dynamical variable $A_0$ is realized.  For example, as can be seen from~(\ref{inf_fraction1}),  high-frequency properties
	 must be more pronounced in higher-order memory functions~\cite{Mokshin/Yulmetyev_book_2006}. Three classes of physical situations can
	 be selected where self-consistency can be realized in their description:
	\begin{enumerate}
		\item[\textbf{A.}]  The set of dynamical variables $\mathbf{A}= \{
		A_0,\; A_1,\; A_2\; \ldots,\; A_{\nu}\}$is initially finite. This situation corresponds to
		a finite set of time scales $\tau_{\nu-1}=\Delta_{\nu}^{-1/2}$:
		\[
		\Delta_1^{-1/2},\; ...,\; \Delta_{\nu-1}^{-1/2} \neq 0,
		\]
		\[
		\Delta_{\nu}^{-1/2}=0.
		\] and is the simplest. In this case, the solutions obtained for the function $\phi_0(t)$, are described by
		undamped harmonic dependences. Some solutions corresponding to this class have been discussed by
		Lee~(see, e.g., \cite{Lee_PRB_1982}). \item[\textbf{B.}]
		The set of dynamical variables $\mathbf{A}= \{ A_0,\; A_1,\; A_2\;
		\ldots,\; A_{\nu},\; \ldots \}$ is characterized by certain relations between
		the time scales
		$\tau_{\nu}=\Delta_{\nu}^{-1/2}$, where $\nu=0$, $1$, $2$, $\ldots$
		and $\Delta_{\nu}^{-1/2}\neq 0$ for any $\nu$. Relations between the
		time scales are ensured by thermodynamic conditions ~\cite{Anatolii_NJP}, with (for example, the phase diagram domain
		under consideration, the size of a characteristic spatial scale, and so on), structure properties of the
		system~\cite{Nigmatullin_PhysA_2006,Nigmatullin_PhysicaA_2000},
		or other reasons~\cite{Wierling_1,Wierling_2}.  In this case, the solutions can relate to very diverse types
		of relaxational (typically, damped) behavior. The possibility of such a realization was first noted by
		Bogoliubov~\cite{Bogolyubov_book,Bogolyubov_1973}, who formulated the postulate of a hierarchy of relaxation times in condensed
		matter. A particular case of equalizing the time scales of different relaxation processes here leads
		to a straightforward realization of Bogoliubov’s idea of a truncated description. The correlation
		approximations proposed by Yulmetyev~\cite{Yulmetyev_P2,Yulmetyev3}, underlie some theoretical models in this class. We
		note that finding the relaxation function is also possible in this class when a relation between time scales in only a group of dynamical variables of the set $\mathbf{A}$ is known. In that case, the general behavior
		of $\phi_0(t)$ can be characterized by complicated (nontrivial) temporal dependences and expressed as a
		functional of the frequency parameters:

		\[
		\phi_{\nu}(t)= \mathcal{F}[\Delta_{\nu},\; \Delta_{\nu+1},\;
		\Delta_{\nu+2},\; \ldots].
		\]
	Specific examples of relaxation processes in this class are the processes associated with microscopic
	one-particle and collective dynamics in equilibrium liquids~\cite{Mokshin/Yulmetyev_book_2006,VNR/NSch_TMF_2004,Ryltsev_2013}.
		\item[\textbf{C.}] Relaxation of various processes associated with dynamical variables in the set $\textbf{A}= \{ A_0,\; A_1,\;
		A_2\; \ldots,\; A_{\nu},\; \ldots \}$, manifests the similarity property
		\begin{equation} \label{eq: scaling}
		\phi_{\nu}(t) \propto \phi_{\mu}(t)^p, \ \ \ \nu,\; \mu =0,\; 1,\;
		2,\; \ldots; \; p>0.
		\end{equation}
		Here, finding $\phi_0(t)$  is also possible in the case where full relaxation of the system occurs on a time
		scale much greater than the one under consideration, and the similarity property of relaxation is
		satisfied only for individual stages. This property allows obtaining solutions in the quasiequilibrium
		and nonequilibrium situations, where system relaxation occurs on a time scale exceeding the one under
		consideration (observed experimentally). A typical example is the structure relaxation in supercooled
		liquids and glasses near vitrification~\cite{Leutheusser_1984},\cite{Gotze},\cite{Kawasaki_1,AVM_FTT_2006,AVM_CP_2007}.
	\end{enumerate}
	
Here, we consider examples corresponding to all three classes.

	\subsection{Finite set of dynamical variables.}
	
We consider the case with a finite set of dynamical variables, i.e., the case where
	$d$ is finite. Such a situation is realized under the condition $A_{\nu} =
	0$ and $\Delta_{\nu} = 0$, and corresponds to nonergodic processes described by an undamped oscillating TCF. \vskip 0.5 cm
	
	\textbf{1. Case $\nu=2$.}
	
	At $\nu=2$,\footnote{The case
		$\nu=1$ is trivial.} we have $A_2=0$ and $\Delta_2=0$. Continued fraction
	(\ref{inf_fraction1})  then transforms
	into the system of two equations
	\begin{equation}
	\left .
	\begin{array}{c}
	1 - s\tilde{\phi}_0(s)  = \Delta_1 \tilde{\phi}_0(s)
	\tilde{\phi}_1(s), \\
	s \tilde{\phi}_1(s)= 1,
	\end{array} \right.
	\end{equation}
	which has the simple solutions
	\begin{equation}
	\phi_0(t)= \cos(\Delta_1^{1/2}t),\label{eq: frr}
	\end{equation}
	\begin{equation}
	\phi_1(t)= 1.
	\end{equation}
	The correlation function of form~(\ref{eq: frr}) reproduces the behavior of an undamped harmonic oscillator. This
	situation is realized, for example, in the case of density fluctuations in a homogeneous electron gas at finite
	wave numbers $k$ and temperature
	$T=0$~\cite{Lee_Physica_Scr_1987}.Another example where such a case is appropriate is the
	dynamics of a chain of classical harmonic oscillators. Here, the time correlation function of the oscillator
	velocity, $\phi_0(t) = \langle
	\vec\upsilon(0) \cdot \vec\upsilon(t) \rangle/\langle
	\vec\upsilon(0) \cdot \vec\upsilon(0) \rangle$, is described by expression~(\ref{eq: frr})~\cite{Florencio_Lee_1985}. \vskip 0.5
	cm
	
	\textbf{2. Case $\nu=3$.}
	
	In this case, the conditions $A_3=0$ and $\Delta_3=0$ hold. Fraction~(\ref{inf_fraction1})  then
	transforms into the system of equations
	\begin{equation}
	\left .
	\begin{array}{c}
	1 - s\tilde{\phi}_0(s)  = \Delta_1 \tilde{\phi}_0(s)
	\tilde{\phi}_1(s), \\
	1 - s\tilde{\phi}_1(s)  = \Delta_2 \tilde{\phi}_1(s)
	\tilde{\phi}_2(s), \\
	s \tilde{\phi}_2(s)= 1,
	\end{array}
	\right.
	\end{equation}
	which can solved by the inverse Laplace transformation. As a result, we obtain the solutions
	\begin{equation} \label{eq: nu_2}
	\phi_0(t)= \frac{1}{\Delta_1 + \Delta_2}\left [ \Delta_2 +\Delta_1
	\cos(\sqrt{\Delta_1+\Delta_2}t) \right ],
	\end{equation}
	\begin{equation}
	\phi_1(t)= \cos(\Delta_2^{1/2}t).
	\end{equation}
	\begin{equation}
	\phi_2(t)= 1.
	\end{equation}
	Expression~(\ref{eq: nu_2})  again corresponds to harmonic behavior of the TCF $\phi_0(t)$, with the oscillation period
	determined by the frequency parameters $\Delta_1$ and $\Delta_2$~\cite{Lee3}.
	
It is obvious from the two cases considered above that an exact analytic solution for the correlation
functions $\phi_{\nu}(t)$
	can be obtained for any finite value of $\nu$.

	\subsection{Infinite set of dynamical variables: \\ Relation between time scales. \label{infin_dimen}}
	
	Various behaviors of $\phi_{\nu}(t)$  are possible in the case where the dynamical variables$\textbf{A}= \{ A_0,\;
	A_1,\; A_2\; \ldots,\; A_{\nu},\; \ldots \}$ form an infinite set.
In this case, the damped character of autocorrelations is universal. \vskip 0.5 cm
	
	\textbf{1. Gaussian relaxation.}
	
	Because the frequency parameters $\Delta_{\nu}$characterize a square relaxation
	scale
	$\tau_{\nu-1}=\Delta_{\nu}^{-1/2}$,, we consider the case where the frequency parameters $\Delta_{\nu}$ are related by an arithmetic
	progression:
	\begin{equation} \label{eq: gaussian_rec}
	\Delta_1, \ \ \Delta_2 = 2 \Delta_1, \ \ \Delta_3 = 3\Delta_1,\ \
	\ldots,\ \  \Delta_{\nu} = \nu\Delta_{1}.
	\end{equation}
	Continued fraction~(\ref{inf_fraction1}) then becomes
	\begin{equation} \label{eq: continued_fraction_Gaussian}
	\tilde{\phi}_0(s) = \frac{1}{\displaystyle s+
		\frac{\Delta_1}{\displaystyle s+ \frac{2\Delta_1}{\displaystyle
				s+\frac{3\Delta_1}{\displaystyle s+\ddots}}}},
	\end{equation}
	which in the temporal representation is the Gaussian function~\cite{Abramowitz}
	\begin{equation} \label{eq: Gaussian}
	\phi_0(t) = \mathrm{e}^{-\Delta_1 t^2/2}.
	\end{equation}
As the best-known example here, we can adduce the relaxation of density fluctuations in an ideal gas and
the relaxation associated with the one-particle dynamics in equilibrium liquids~\cite{Mokshin/Khusnutdinoff/Yulmetyev_2006,Hansen/McDonald_book_2006}.
	
On the other hand, the exact correspondence between the frequency parameters established by relations ~(\ref{eq:
		gaussian_rec}), suggests the possibility of a quantitative estimate of the deviation of the relaxation process from
	the Gaussian dependence by simple comparison of the parameters $\Delta_{\nu}$:
	\begin{equation} \label{eq: non-Gaussian}
	\alpha_{n} = \frac{n}{n+1}\frac{\Delta_{\nu+1}}{\Delta_{\nu}}-1.
	\end{equation}
In the case of Gaussian relaxation, we have  $\alpha_{n}=0$, while deviations of the $\alpha_{n}$ from zero characterize
non-Gaussian behavior of
	$\phi_0(t)$~\cite{AVM_DNC_2013}. \vskip 0.5cm
	
	\textbf{2. Damped oscillating correlator.}
	
We consider a particular case where the frequency parameters are finite and take coincident values, which is associated with equalizing the characteristic time scales:
	\begin{equation} \label{eq: condition_for_Bessel}
	\Delta_1^{-1/2} = \Delta_2^{-1/2} = \Delta_3^{-1/2} = \ldots =
	\Delta_{\nu}^{-1/2}.
	\end{equation}
Continued fraction~(\ref{inf_fraction1}) then becomes
	\begin{equation} \label{eq: continued_fraction_Bessel}
	\widetilde{\phi}_0(s) = \frac{1}{\displaystyle s+
		\frac{\Delta_1}{\displaystyle s+ \frac{\Delta_1}{\displaystyle
				s+\ddots}}}.
	\end{equation}
	\begin{equation} \label{eq: Bessel_freq}
	\widetilde{\phi}_0(s)=\frac{-s+\sqrt{s^2+4\Delta_1}}{2\Delta_1}.
	\end{equation}
	Applying the inverse Laplace transformation to expression~(\ref{eq:
		Bessel_freq}), we obtain the TCF
	\begin{equation} \label{eq: Bessel_time}
	\phi_0(t) = \frac{1}{\Delta_1^{1/2}t}J_{1}(2\Delta_1^{1/2}t),
	\end{equation}
	where $J_1(\ldots)$ is the Bessel function of the first kind. Such a relaxation occurs in processes with damped
	harmonic behavior. For example, expression~(\ref{eq:
		Bessel_time})is an exact form of the autocorrelation function of the
	velocity of an impurity particle in a linear chain of identical harmonic oscillators~\cite{Rubin,Mokshin/Yulmetyev,Zwanzig_book_2001}, where
	$\Delta_1 = 2K/M$ and $K$ is the coupling coefficient for neighboring particles (spring constant), and $M$ is the mass
	of the impurity particle. Relaxation of the two-dimensional electron gas density at the temperature
	$T=0$ and finite wave numbers $k$ is also described by expression~(\ref{eq: Bessel_time})
	(see~\cite{Lee_Physica_Scr_1987}).
	
The above examples clearly demonstrate that with the known relation between the corresponding
relaxation scales $\tau_{\nu-1}=\Delta_{\nu}^{-1/2}$ we can find the original TCF $\phi_0(t)$, exactly and assess the properties of its
frequency representation $\widetilde{\phi}_0(s)$. \vskip 0.5cm
	
	\textbf{3. Kinetic coefficients: Self-diffusion and the density of vibrational states in an equilibrium simple liquid.}
	
The kinetic coefficients (diffusion coefficient $D$,  shear viscosity $\eta$,
	 and heat conduction coefficient $\lambda_T$) in one-component liquids
	$\mathcal{P}=\{D,\; \eta,\; \lambda_T \}$
	are related to correlation functions of the
	flux variables ${A}_0=\{\vec{v},\;P_{xy},\; J_{0}^{ez} \}$
	by Kubo–Green’s integral relations~\cite{Hansen/McDonald_book_2006} of the general form:
	\begin{equation} \label{eq: Green/Kubo_gen}
	\mathcal{P} = \mathcal{Q} \int_{0}^{\infty} \frac{\langle {A}_0(0)
		{A}_0(t) \rangle}{ \langle {A}_0(0)^2 \rangle} dt,
	\end{equation}
	where the
 $\mathcal{Q}$ factor is
	\begin{equation}
	\mathcal{Q} = \left \{ \frac{k_B T}{m},\; \frac{(P_{xy})^2}{k_B T
		V}, \; \frac{(J_0^{ez})^2}{k_B T^2 V} \right \} .
	\end{equation}
Components of the pressure tensor are determined from the virial formula~\cite{Evans/Morriss_Stat_Mech_2008}:
	\begin{equation}
	P_{\alpha, \beta} = \sum_{i=1}^{N} \left ( m v_{i\alpha}v_{i\beta}
	+ \frac{1}{2} \sum_{i\neq j}^{N} F_{ij\alpha} r_{ij\beta} \right	), \; \; \alpha,\beta =x,y,z,
	\end{equation}
	where $F_{ij\alpha}$ denotes the $\alpha$ component of the force acting between particles $i$ and $j$, separated by a distance
	$r_{ij}$. The heat flux can be evaluated from the expression
	\begin{equation}
	J_0^{ez} = \sum_{i=1}^{N} v_{iz} \left ( \frac{m
		|\textbf{v}_i|^2}{2} + \frac{1}{2}\sum_{i\neq j}^N U(r_{ij})
	\right ) - \frac{1}{2} \sum_{i=1}^{N} \sum_{i\neq j}^N
	\textbf{v}_i \textbf{r}_{ij}\frac{\partial U(r_{ij})}{\partial
		z_{ij}},
	\end{equation}
	where $U(r_{ij})$ is the interparticle interaction potential \cite{AVM_DNC_2013}, \cite{AVM_FTT_2011}.

	On the other hand, recalling that $\tilde{\phi}_0^{AA}(s)$ is the Laplace transform of the TCF $\phi_0^{AA}(t)=\langle {A}_0(0) {A}_0(t)
	\rangle/\langle {A}_0(0)^2 \rangle $, we can write \Ref{eq:
		Green/Kubo_gen} as
	\begin{equation}
	\mathcal{P} = \mathcal{Q} \lim_{s \to 0} \tilde{\phi}_0^{AA}(s),
	\end{equation}
	where
	\begin{equation}
	\widetilde{\phi}_{0}^{AA}(s)= \frac{1}{\displaystyle
		s+\frac{\Delta_{1}^{AA}}{\displaystyle
			s+\frac{\Delta_{2}^{AA}}{\displaystyle
				s+\frac{\Delta_{3}^{AA}}{\displaystyle s+\ddots}}}}. \label{eq:
		cont_fraction}
	\end{equation}
	
In accordance with the self-consistent approach, we assume that at some $\nu$-th level of the relaxation
hierarchy, the time scales
	$\tau_{\nu-1}^{AA}=1/\sqrt{\Delta_{\nu}^{AA}}$ !!! equalize:

	\begin{equation}
	\frac{\tau_{\nu}^{AA}}{\tau_{\nu-1}^{AA}} \to 1, \ \ \ \nu=1,\;
	2,\;3,\; \ldots .
	\end{equation}
	Then the kinetic coefficients can be expressed in terms of the eigenfrequency parameters:
	\begin{equation} \label{eq: Transport_gen}
	\mathcal{P} =  \mathcal{Q} \times \left\{
	\begin{array}{rl}
	\displaystyle{\frac{\Delta_2^{AA}\ldots \Delta_{\nu-1}^{AA}}{\Delta_1^{AA}\ldots {\sqrt{\Delta_{\nu}^{AA}}}}}, & \nu=1,\;3,\;5,\;\ldots \\
	\displaystyle{\frac{\Delta_2^{AA}\ldots \sqrt{\Delta_{\nu}^{AA}}}{\Delta_1^{AA}\ldots \Delta_{\nu-1}^{AA}}}, & \nu=2,\;4,\;6,\;\ldots \\
	\end{array}
	\right. ,
	\end{equation}
which are to be calculated for each process in accordance with general definition~\Ref{eq: freq_par}. \vskip 0.5 cm

	As an example, we consider finding the self-diffusion coefficient in a liquid inside a volume  $V$, with $N$ particles interacting via a spherical potential $U(r)$. In that case, the quantity
	\begin{equation}
	\phi_0^{vv}(t) = \frac{\langle \vec{v}(0) \cdot \vec{v}(t)
		\rangle}{\langle \vec{v}(0) \cdot \vec{v}(0) \rangle}
	\end{equation}
	is the particle velocity autocorrelation function, and
	$\Delta_j^{vv}$, $j=1,\; 2,\; \ldots$, are the corresponding frequency parameters. The spectral density of the velocity autocorrelation function yields the density $G(\omega)$,of vibrational
	states of the multiparticle system,
	\begin{equation} \label{eq: VDOS_1}
	G(\omega) = \int \mathrm{e}^{i\omega t} \phi_0^{vv}(t) dt,
	\end{equation}
	\begin{equation} \label{eq: VDOS_2}
	G(\omega) \sim \mathrm{Re} \; \widetilde{\phi}_{0}^{vv}(s),\ \ s=
	\sigma+i\omega.
	\end{equation}

 From~\Ref{eq: Transport_gen} at $\nu=1$, we obtain
	\begin{equation}\label{first_diff}
	D=\frac{k_{B}T}{m}  \frac{1}{\sqrt{\Delta_1^{vv}}},
	\end{equation}
	where
	\begin{equation}
	\Delta_1^{vv} = \frac{\langle \dot{\vec{v}} \cdot \dot{\vec{v}}
		\rangle }{\langle \vec{v} \cdot \vec{v} \rangle } = \frac{4\pi
		n}{3}\int_{0}^{\infty}dr~ g(r) r^{3} \left \lbrack
	\frac{3}{r^2}\frac{\partial U(r)}{\partial
		r}+\frac{\partial}{\partial r}\left(\frac{\partial U(r)}{r\partial
		r}\right)\right \rbrack ,
	\end{equation}
and	$g(r)$ is the radial distribution function of particles. The spectral density then becomes
	\begin{equation}
	G(\omega) \sim \frac{(4 \Delta_1^{vv} -
		\omega^2)^{1/2}}{2\Delta_1^{vv}}.
	\end{equation}
	
	At $\nu=2$, we find
	\begin{equation}\label{second_diff}
	D =\frac{k_{B}T}{m}\frac{\sqrt{\Delta_2^{vv}}}{\Delta_1^{vv}},
	\end{equation}
	where
	\begin{eqnarray}
	\Delta_2^{vv}&=&\frac{8\pi n}{3m}\int_{0}^{\infty}dr~ g(r)\left [
	3\left (\frac{\partial U(r)}{\partial r}\right )^{2} \right.
	+\left (r\frac{\partial}{\partial r}\left(\frac{\partial
		U(r)}{r\partial r}\right)\right )^{2}+ \left. \frac{\partial
		U(r)}{\partial r} \frac{\partial}{\partial r}\left(\frac{\partial
		U(r)}{r\partial r}\right)
	\right ]+\nonumber\\
	&+&\frac{8\pi^{2}n^{2}}{3m}\int_{0}^{\infty} \int_{0}^{\infty}dr
	dr_{1}~ r^{2} r_{1}^{2}\int_{-1}^{1}d\beta_r ~
	g_{3}(\vec{r},\vec{r}_{1}) \left [ \frac{3}{rr_{1}} \frac{\partial
		U(r)}{\partial r}
	\frac{\partial U(r_{1})}{\partial r_{1}}+ \right. \nonumber\\
	& &+\frac{r}{r_{1}} \frac{\partial U(r_{1})}{\partial r_{1}}
	\frac{\partial}{\partial r}\left(\frac{\partial U(r)}{r\partial
		r}\right) +\frac{r_{1}}{r} \frac{\partial U(r)}{\partial r}
	\frac{\partial}{\partial r_{1}}\left(\frac{\partial
		U(r_{1})}{r_{1}\partial r_{1}}\right)+ \nonumber\\
	& &+ \left. rr_{1}\frac{\partial}{\partial
		r_{1}}\left(\frac{\partial U(r_{1})}{r_{1}\partial r_{1}}\right)
	\frac{\partial}{\partial r}\left(\frac{\partial U(r)}{r\partial
		r}\right)\beta_r^{2} \right ]. \nonumber
	\end{eqnarray}
	Here $g_{3}(\vec{r},\vec{r}_{1})$ is the distribution function of three particles, and $\beta_r$ is the cosine of the angle between the
	vectors
	$\vec{r}$ and $\vec{r}_{1}$. We then find the density of vibrational states
	\begin{equation}
	G(\omega) \sim \frac{2 \Delta_1 \Delta_2 \sqrt{4\Delta_2 -
			\omega^2}}{\omega^2(2\Delta_2 - \Delta_1)^2 +\Delta_1^2(4\Delta_2
		- \omega^2)}
	\end{equation}
	
	At $\nu=3$, we obtain
	\begin{equation}
	D=\frac{k_{B}T}{m}\frac{\Delta_2^{vv}}{\Delta_1^{vv}\sqrt{\Delta_3^{vv}}},
	\label{third_diff}
	\end{equation}
where the frequency  $\Delta_3^{vv}$ contains two-, three-, and four-particle correlation functions
	\cite{Bansal}:
	\begin{eqnarray}
	\Delta_3^{vv}&=&\frac{4 \pi}{m^3} \int d\vec{r} ~g(\vec{r})
	[U_{xy}(r)U_{xx}(r)U_{yy}(r) +  3 k_B T U_{xyz}^2(r)]+ \nonumber\\
	& &\ \ \ \ \ +\frac{n^2}{m^3} \int \int d\vec{r}d\vec{r}_1 ~
	g_3(\vec{r},\vec{r}_1)[3k_B T
	U_{xxy}(r)U_{xxy}(r_1)]+\nonumber\\
	& &+U_{xx}(r)(6U_{xy}(r)U_{xy}(r_1)-
	U_{xy}(r)U_{xy}(|\vec{r}-\vec{r}_1|))]+ \nonumber\\
	& &+\frac{n^3}{m^3}\int\int\int d\vec{r} d\vec{r}_1 d\vec{r}'_2 ~
	g_4(\vec{r},\vec{r}_1,\vec{r}_2)U_{xx}(r)U_{xy}(r_1)U_{xy}(r'_2),\nonumber
	\end{eqnarray}
	\[
	U_{xyz}(r)=\frac{\partial^3U(r)}{\partial r_x
		\partial r_y
		\partial r_z}. \nonumber
	\]
	The density of vibrational states is here given by
	\begin{equation}
	G(\omega) \sim \frac{2 \Delta_1 \Delta_2 \Delta_3 \sqrt{4\Delta_3
			- \omega^2}}{[2\Delta_1 \Delta_3 - \omega^2(2\Delta_3 -
		\Delta_2)]^2 + \omega^2 \Delta^2 (4\Delta_3 - \omega^2)}.
	\end{equation}

	%
	%
	
	\vskip 0.5cm
	
	\textbf{4. Density fluctuations in equilibrium simple liquids.}
	
	Collective dynamics of particles in a
	liquid can be quantitatively characterized by the experimentally measured quantity, the dynamical structure
	factor $S(k,\omega)$, which is the Fourier image of the scattering function~\cite{Hansen/McDonald_book_2006,Novikov_2011_2,Novikov_2011_1,Novikov_2012}:
	\begin{equation} \label{eq: dynamic_str_factor}
	S(k,\omega) = \frac{S(k)}{2\pi} \int_{-\infty}^{\infty}
	\mathrm{e}^{-i\omega t} F(k,t)dt,
	\end{equation}
	where $S(k)$  is the static structure factor. In turn, the scattering function $F(k,t)$ is the TCF of the local
	density $N$  fluctuations for particles in the liquid:
	\begin{equation}
	F(k,t)=\frac{\langle \delta \rho_{k}^{*}(0)
		\delta \rho_{k}(t) \rangle}{\langle
		\mid \delta \rho_{k}(0)\mid^{2} \rangle},  \label{TCF}
	\end{equation}
	where $k=|\vec{k}|$ is the wave number. Therefore, it is quite convenient to choose the original dynamical variable
 $A_0$
	as~\cite{Mokshin/Yulmetyev_PRE_2001,Mokshin/Yulmetyev_JPCM_2003,AVM_1,AVM_2,AVM_3}
	\begin{equation}
	A_0 = \delta \rho(k)=\frac{1}{\sqrt{N}}
	\sum_{j=1}^{N}\textrm{e}^{i \vec{k} \cdot
		\vec{r}_{j}}-\delta_{\vec{k},0}n, \label{fluct.}
	\end{equation}
where $\vec{r}_{j}(t)$  is the radius vector describing the position of the~$j$-th particle at an instant $t$. With recurrence
relation~\Ref{RR1} we obtain an infinite set of orthogonal dynamical variables~\cite{Mokshin/Khusnutdinoff/Yulmetyev_2006}
	\begin{eqnarray}
	\textbf{A}(k)&=&\{A_0(k),\; A_1(k),\; A_2(k),\; \ldots ,\; A_j(k),
	\; \ldots\},\ \ \ 
	\\
	& &\langle A_j^*A_l \rangle = \delta_{j,l}\langle
	|A_j|^2 \rangle,\nonumber\\
	& &A_0(k)\equiv \delta \rho(k) \nonumber
	\end{eqnarray}
	with TCFs of the form
	\begin{equation}
	\label{TCFs} M_j(k,t)=\frac{\langle
		A_j^{*}(k,0)A_j(k,t)\rangle}{\langle | A_j(k,0)|^{2} \rangle},
	\end{equation}
	\[
	M_0(k,t) \equiv  F(k,t),
	\]
	which have properties~(\ref{prop_TCF}).
	
	We define normalized frequency moments of the dynamical structure factor:
	\begin{eqnarray}
	\omega^{(2p)}(k)= \frac{\int_{-\infty}^{\infty}\omega^{2p}
		S(k,\omega)d\omega}{\int_{-\infty}^{\infty}S(k,\omega)d\omega}=
	\left. (-i)^p\frac{d^{p}F(k,t)}{dt^p} \right |_{t=0}, \label{fr_m}
	\end{eqnarray}
	\[
	p=1,\ 2,\ \ldots \ .
	\]
	
	Only even moments ($p=2,$ $4$,
	$\ldots$), take finite values, and odd moments vanish. From \Ref{RR1}, using the
fact that the relation
	\begin{equation}
	\left. i^{(2j)}\frac{d^{2j} \phi(t)}{dt^{2j}}\right
	|_{t=0}=\frac{\langle [(i \hat{\mathcal{L}} A_{0})^j]^* (i
		\hat{\mathcal{L}} A_{0})^j \rangle}{\langle|A_{0}|^2\rangle},
	\end{equation}
	holds for a fixed wave number $k$, we can obtain expressions that relate frequency moments $\omega^{(2j)}(k)$ to the frequency parameters
	$\Delta_j(k)$:
	\begin{eqnarray} \label{Eq_12}
	\Delta_{1}(k)&=&\omega^{(2)}(k),
	\\
	\Delta_{2}(k)&=&\frac{\omega^{(4)}(k)}{\omega^{(2)}(k)}-\omega^{(2)}(k),
	\nonumber\\
	\Delta_{3}(k)&=&\frac{\omega^{(6)}(k)\omega^{(2)}(k)-(\omega^{(4)}(k))^{2}}
	{\omega^{(4)}(k)\omega^{(2)}(k)-(\omega^{(2)}(k))^{3}},
	\nonumber \\
	\Delta_{4}(k)&=&\frac{1}{\Delta_{1}(k)\Delta_{2}(k)\Delta_{3}(k)}
	\Bigg\{\omega^{(8)}(k)-\Delta_{1}(k)\nonumber\\
	& &\times \Biggl[
	\Bigl(\Delta_{1}(k)+\Delta_{2}(k)\Bigr)^{3}+2\Delta_{2}(k)\Delta_{3}(k)
	\Bigl(\Delta_{1}(k) +\Delta_{2}(k) \Bigr)\nonumber\\
	& &\hskip 10cm +\Delta_{2}(k)\Delta_{3}^{2}(k)
	\Biggr]\Bigg\}, \nonumber\\
	\Delta_{5}(k)&=&\frac{1}{\Delta_{1}(k)\Delta_{2}(k)\Delta_{3}(k)\Delta_{4}(k)}\Bigg\{\omega^{(10)}(k)-2\omega^{(8)}(k)
	\Big[\Delta_{1}(k)+\Delta_{2}(k)+\nonumber\\
	& &\hskip 10cm \Delta_{3}(k)+\Delta_{4}(k)\Big]+ \nonumber\\
	& &
	+\omega^{(6)}(k)\Big[\Delta_{1}^{2}(k)+2\Delta_{1}(k)\Delta_{2}(k)+\Delta_{2}^{2}(k)+4\Delta_{1}(k)\Delta_{3}(k)
	\nonumber \\& &
	+2\Delta_{2}(k)\Delta_{3}(k)+\Delta_{3}^{2}(k)+4\Delta_{1}(k)\Delta_{4}(k)
	\Big]\Biggl\}+\Delta_{4}(k). \nonumber
	\end{eqnarray}
Expressions of this form are also known as sum rules~\cite{Temperly2}.
	
	Expressions for the frequency parameters $\Delta_j(k)$ follow from definition \Ref{eq: freq_par} with Eqs. \Ref{RR1} and
	\Ref{fluct.}  taken into account.
	 For example, for the first three frequency parameters
	$\Delta_1(k)$, $\Delta_2(k)$ and $\Delta_3(k)$
	we have~\cite{Mokshin/Khusnutdinoff/Yulmetyev_2006}:
	\begin{eqnarray}
\Delta_{1}(k)=\frac{k_BT}{m}\frac{k^2}{S(k)},
	\nonumber \\
	\Delta_{2}(k)=  3\frac{k_BT}{m}k^2
	+ \frac{\rho}{m} \int \nabla_l^2 U(r)[1-
	\exp(i\vec{k}\cdot\vec{r})]g(r) d^3r - \Delta_1(k),
	\\
	\Delta_{3}^2(k)= \frac{15}{\Delta_{2}(k)}	\left (
	\frac{k_BT}{m}k^2 \right )^2 
	-\frac{\left [\Delta_{1}^2(k)+\Delta_{2}^2(k) \right
		]^2}{\Delta_{2}^2(k)} +
	\frac{\mathcal{U}(k)}{\Delta_1^2(k)\Delta_2^2(k)},\nonumber
	\label{fr_c}
	\end{eqnarray}

	where $\mathcal{U}(k)$  denotes a combination of integral expressions that contains the interparticle interaction potential
	$U(r)$ and the three-particle distribution function
	$g_{3}(\vec{r},\vec{r}')$. The full expression for $\mathcal{U}(k)$
	is given in \cite{Michler_JPF_1977} (see. p.~870). We
	note that in addition to a considerable complication of the analytic expressions for frequency relaxation
	parameters $\Delta_j^2(k)$ with the increasing order $j$,  they also start involving equilibrium distribution functions
	of $j$-particle groups:
	\begin{equation}
	\Delta_j(k) = \mathcal{F}[\Delta_1(k),
	\Delta_2(k),\ldots,\Delta_{j-1}(k); g(r),g_3(\bar{r}),\ldots,
	g_j(\bar{r})].
	\end{equation}
	
	With the density fluctuations $A_{0}(k)=\delta \rho(k)$chosen as the original dynamical variable [see (\ref{fluct.})]we
	can express $A_1(k)$ for $k\neq 0$ as
	\begin{equation} A_{1}(k)=
	\frac{1}{mV} \sum_{j=1}^{N}(m_j v_j^l) \textrm{e}^{i
		\vec{k}\cdot\vec{r}_{j}}, \label{A_1}
	\end{equation}
	where the index $l$ denotes the longitudinal component (parallel to
	$\vec{k}$). For $A_2(k)$ and $A_3(k)$ we find
	\begin{eqnarray}
	A_{2}(k)&=&\frac{1}{mV} \sum_{j=1}^{N} \left \lbrace
	\frac{(m_{j}v_{j}^{l})^{2}}{m}+i \sum_{i>j=1}^{N}
	\vec{\nabla}_{j}u(j,i)\vec{k} \lbrack1-\textrm{e}^{i \vec{k}\cdot
		(\vec{r}_{i} -\vec{r}_{j})} \rbrack \right \rbrace \textrm{e}^{i
		\vec{k} \cdot
		\vec{r}_{j}}- \nonumber\\
	&-& \Omega_{1}^{2}(k) \frac{1}{V} \sum_{j=1}^{N} \textrm{e}^{i
		\vec{k} \cdot \vec{r}_{j}}, \label{A_2}
	\end{eqnarray} \noindent
	\begin{eqnarray}
	A_{3}(k) &=& \frac{1}{V}
	\sum_{j=1}^{N}\frac{(\vec{p}_{j}\vec{k})^{3}}{m^{3}}\textrm{e}^{i
		\vec{k} \cdot \vec{r}_{j}}+  \frac{2i }{Vm^{2}}\sum_{i>j=1}^{N}
	(\vec{k}\vec{\nabla}_{j})u(i,j) \left \lbrace \right. (\vec{p}_{j}
	\vec{k})\textrm{e}^{i \vec{k} \cdot
		\vec{r}_{j}}-(\vec{p}_{i}\vec{k})\textrm{e}^{i  \vec{k} \cdot
		\vec{r}_{i}} \rbrace +
	\nonumber\\
	&+& \sum_{i>j=1}^{N} \frac{\vec{p}_{j}
		\vec{\nabla}_{j}}{Vm^{2}}(\vec{k}\vec{\nabla}_{j}) u(j,i) \left
	\lbrace \textrm{e}^{i \vec{k} \cdot \vec{r}_{j}}-\textrm{e}^{i
		\vec{k} \cdot \vec{r}_{i}} \right \rbrace
	-[\Delta_{1}(k)+\Delta_{2}(k)]A_{1}(k). \label{A_3}
	\end{eqnarray}
	The dynamical variable $A_4(k)$ is
	\begin{eqnarray} A_{4}(k)&=&\frac{1}{V}
	\sum_{j=1}^{N}\frac{(\vec{p}_{j} \vec{k})^{4}}{m^{4}}\textrm{e}^{i
		\vec{k} \cdot \vec{r}_{j}}+  \frac{3i
	}{Vm^{3}}\sum_{i>j=1}^{N}(\vec{k} \vec{\nabla}_{j})u(j,i) \lbrack
	(\vec{p}_{j}\vec{k})^{2}\textrm{e}^{i \vec{k} \cdot \vec{r}_{j}}-
	(\vec{p}_{i}\vec{k})^{2}\textrm{e}^{i \vec{k} \cdot \vec{r}_{i}}
	\rbrack +
	\nonumber\\
	&+&\frac{2}{Vm^{3}}\sum_{i>j=1}^{N}\vec{p}_{j}\vec{\nabla}_{j}(\vec{k}
	\vec{\nabla}_{j})u(j ,i) \left \lbrack
	(\vec{p}_{j}\vec{k})\textrm{e}^{i \vec{k} \cdot
		\vec{r}_{j}}-(\vec{p}_{i}\vec{k})\textrm{e}^{i \vec{k} \cdot
		\vec{r}_{i}} \right \rbrack -
	\nonumber\\
	&-&\frac{2}{Vm^{2}}\sum_{i>j=1}^{N}(\vec{k}
	\vec{\nabla}_{j})^{2}u^{2}(j,i) \left \lbrack \textrm{e}^{i
		\vec{k} \cdot \vec{r}_{j}}+\textrm{e}^{i \vec{k} \cdot
		\vec{r}_{i}} \right \rbrack - \nonumber\\
	&-& \frac{i
	}{Vm^{3}}\sum_{i>j=1}^{N}\vec{p}_{j}^{2}\vec{\nabla}_{j}^{2}
	(\vec{k} \vec{\nabla}_{j})
	u(j,i) \left \lbrack \textrm{e}^{i \vec{k} \cdot
		\vec{r}_{j}}-\textrm{e}^{i \vec{k} \cdot \vec{r}_{i}} \right
	\rbrack + \nonumber\\ &+& \frac{i
	}{Vm^{2}}\sum_{i>j=1}^{N}\vec{\nabla}_{j}u(j,i)(\vec{\nabla}_{j}-\vec{\nabla}_{i})
	(\vec{k}\vec{\nabla}_{j})u(j,i)\left \lbrack \textrm{e}^{i \vec{k}
		\cdot \vec{r}_{j}}-\textrm{e}^{i \vec{k} \cdot \vec{r}_{i}} \right
	\rbrack - \nonumber\\
	&-& [\Delta_{1}(k)+\Delta_{2}(k)+\Delta_{3}(k)]A_{2}(k)-
	[\Delta_{1}(k)+\Delta_{2}(k)]\Delta_{1}^{2}(k) A_{0}(k).
	\label{A_4}
	\end{eqnarray}
	
	We see from expressions  \Ref{fluct.}, \Ref{A_1}, \Ref{A_2}, \Ref{A_3},
	\Ref{A_4}  that with the local density fluctuation $A_0(k)$  chosen as
	the original dynamical variable, the function
	$M_1(k,t)$ is the TCF of the longitudinal momentum component
	fluctuations, $M_2(k,t)$ is directly related to the TCF of energy fluctuations, and so on. Therefore, in the
	limit of small wave numbers $(k
	\to 0)$,$M_0(k,t)$, $M_1(k,t)$ and $M_2(k,t)$can be associated with the autocorrelators corresponding to three conserved hydrodynamic variables. The higher-order dynamical variables ($A_j$ at $j\geq 3$)  describe relaxation of more "complicated" processes, which contain crosscorrelations of the
	momenta, energies, fluxes, and so on. For example, the process characterized by
	$A_3(k)$, can be identified
	with fluctuations of the longitudinal energy flux~\cite{Mokshin/Yulmetyev_book_2006}.
	
The description can be reduced in the framework of the postulate of equalizing time scales
	\[
	\frac{\tau_{j}}{\tau_{j-1}} \to 1, \ \ \ j=1,\ 2,\ \ldots
	\]
	In the simplest approximation that agrees with provisions of hydrodynamic theory, we assume that the
	relaxation times of the subsequent TCFs are comparable, in contrast to the scales of the first three dynamical variables $A_0$, $A_1$ and $A_2$:
	$\tau_3(k) \simeq \tau_4(k) \simeq \tau_j(k),\; j>3$,  which is
	\begin{equation}
	\Delta_4(k)=\Delta_5(k)=\Delta_6(k)= \ldots=\Delta_j(k), \ \ \
	j\geq4. \label{cond}
	\end{equation}
	Similarly to (\ref{inf_fraction1}) from (\ref{eq:
		continued_fraction_Bessel})-(\ref{eq: Bessel_time}) we then obtain
	\begin{equation} \label{eq: M3_Bessel_freq}
	\widetilde{M}_3(s)=\frac{-s+\sqrt{s^2+4\Delta_4}}{2\Delta_4}
	\end{equation}
	and
	\begin{equation} \label{eq: M3_Bessel_time}
	M_3(t) = \frac{1}{\Delta_4^{1/2}t}J_{1}(2\Delta_4^{1/2}t),
	\end{equation}
	Substituting (\ref{eq: M3_Bessel_freq}) in (\ref{inf_fraction1}) and taking (\ref{eq: dynamic_str_factor})  into account, we find the dynamical structure factor in the form

	\begin{eqnarray} \label{Basic} S(k, \omega)&=& \frac{S(k)}{2 \pi}
	\Delta_{1}(k) \Delta_{2}(k) \Delta_{3}(k) [4 \Delta_{4}(k)-
	\omega^{2}]^{ \frac{1}{2}} \lbrace \Delta_{1}^{2}(k)
	\Delta_{3}^{2}(k)+
	\nonumber \\
	&+&\omega^{2}[\Delta_{1}^{2}(k) \Delta_{4}(k)-2 \Delta_{1}(k)
	\Delta_{3}^{2}(k)- \Delta_{1}^{2}(k)
	\Delta_{3}(k) +\nonumber\\
	& &\ \ \ \ \ +2 \Delta_{1}(k) \Delta_{2}(k) \Delta_{4}(k)-
	\Delta_{1}(k) \Delta_{2}(k) \Delta_{3}(k)+ \Delta_{2}^{2}(k)
	\Delta_{4}(k)]+
	\nonumber \\
	&+&\omega^{4}[ \Delta_{3}^{2}(k)-2 \Delta_{1}(k) \Delta_{4}(k) +2
	\Delta_{1}(k) \Delta_{3}(k) -2 \Delta_{2}(k) \Delta_{4}(k)+
	\nonumber\\
	& &\ \ \ \ \ \ \ \ \ \ \ \ \ \ \ \ \ \ \ \ \ \ \ \ \ \ \ \ \ \ \ \
	\ \ \ \ \ \ \ \ \ \ \ \ \ \ \ \ \ \ \ \ \ \ \ \ \ \ \ \ +
	\Delta_{2}(k) \Delta_{3}(k)]+
	\nonumber\\
	&+& \omega^{6}[ \Delta_{4}(k)- \Delta_{3}(k)] \rbrace^{-1}.
	\end{eqnarray} As can be seen from $S(k,\omega)$ in the domain of finite k corresponding to microscopic spatial scales
	is completely defined by the first four frequency parameters, which contain information about microscopic
	properties of the system and are expressed in terms of the interparticle interaction potential, and also two-,
	three-, and four-particle distribution functions
	\cite{Mokshin/Yulmetyev_book_2006,Gufan,Gufan_2}.
	
It is known that the dynamical structure factor spectra recorded in experiments on inelastic scattering
of neutrons and X-rays in simple liquids involve not only the elastic component (at zero frequency) but also
two symmetric inelastic peaks at the frequencies $\pm \omega_c \neq 0$ (see,
	e.g., \cite{Scopigno_RMP_2005}).  Although the overall shape of
	the spectrum
	$S(k,\omega)$ at finite values of the wave number resembles the shape of the known hydrodynamic
	triplet observed in experiments on light scattering, the peaks are not distinctly separated in this case, in
	contrast to the hydrodynamic case~\cite{Tokarchuk_TMF_2008,Wax_2013,AVM_JETP_2009,AVM_Izv_RAN_2010,AVM_JPCS_2012,AVM_Surf_2014}.
	 Inelastic properties of scattering spectra are manifested in the
	longitudinal flux TCF~\cite{Mokshin_JPCM_2006}
	\begin{equation}
	G_J(k,t) = \frac{\langle J^L(k,0)J^L(k,t) \rangle}{\langle
		|J^L(k,0)^2| \rangle},
	\end{equation}
	which is related to the dynamical structure factor as
	\begin{eqnarray} \label{Skw_current}
	S(k) \Delta_1(k) \widetilde{G}_J(k,\omega)= \omega^2 S(k,\omega).
	\end{eqnarray}
	In turn, relation~\Ref{Skw_current} can be obtained from the equality
	\begin{eqnarray}
	\Delta_1(k) G_J(k,t) = -\frac{\partial^2 F(k,t)}{\partial t^2}.
	\label{cur_rel}
	\end{eqnarray}
	
	For one-component simple liquids,
	$\widetilde{G}_J(k,\omega)$ has a minimum at the zero frequency ($\omega=0$), and two
	high-frequency maxima (at $\omega \neq 0$).  The position and width of the maxima in the spectrum of $\widetilde{G}_J(k,\omega)$
	 are
	determined by solutions of the so-called dispersion equation
	\begin{eqnarray}
	s+\frac{\Delta_1(k)}{s}+\Delta_2(k)\widetilde{M}_{2}(k,s)=0
	\label{condd}
	\end{eqnarray}
	for $s=s(k)$, where the expression for $\widetilde{M}_{2}(k,s)$
	follows from (\ref{eq: M3_Bessel_freq}) in the form
	\begin{eqnarray}
	\widetilde{M}_{2}(k,s)=\frac{2\Delta_4(k)}{s[2\Delta_4(k)-
		\Delta_3(k)] +\Delta_3(k)\sqrt{s^2+ 4\Delta_4(k)}}. \nonumber \\
	\label{a_2}
	\end{eqnarray}
	In the general form, Eq.~(\ref{condd}) has complex solutions $s=\textrm{Re}[s(k)]+i\textrm{Im}[s(k)]$, where
	$\textrm{Im}[s(k)]$  determines
	the position of inelastic peaks in
	$\widetilde{G}_J(k,\omega)$ and $\textrm{Re}[s(k)]$ characterizes the width of these peaks.
	
	We introduce the dimensionless quantities
	\begin{eqnarray}
	\mathcal{Q}(k)=2\frac{\Delta_4(k)}{\Delta_3(k)} - 1, \label{Q}
	\end{eqnarray}
	\begin{eqnarray}
	\xi(k)=\frac{s^2}{\Delta_4(k)}. \label{small_p}
	\end{eqnarray}
	The condition for the existence of high-frequency peaks in Eq.~(\ref{condd}) can then be written as
	\begin{eqnarray}
	s^2+s\Delta_2(k)\frac{1 +\mathcal{Q}(k)}{s\mathcal{Q}(k)
		+\sqrt{s^2+ 4\Delta_4(k)}} +\Delta_1(k)=0. \label{cond_fin}
	\end{eqnarray}
To analyze this equation, we consider the following limit situations: \vskip 0.5 cm
	
\begin{figure} [hbt!]
		\vskip -0.5cm
		\centerline{\epsfig{figure=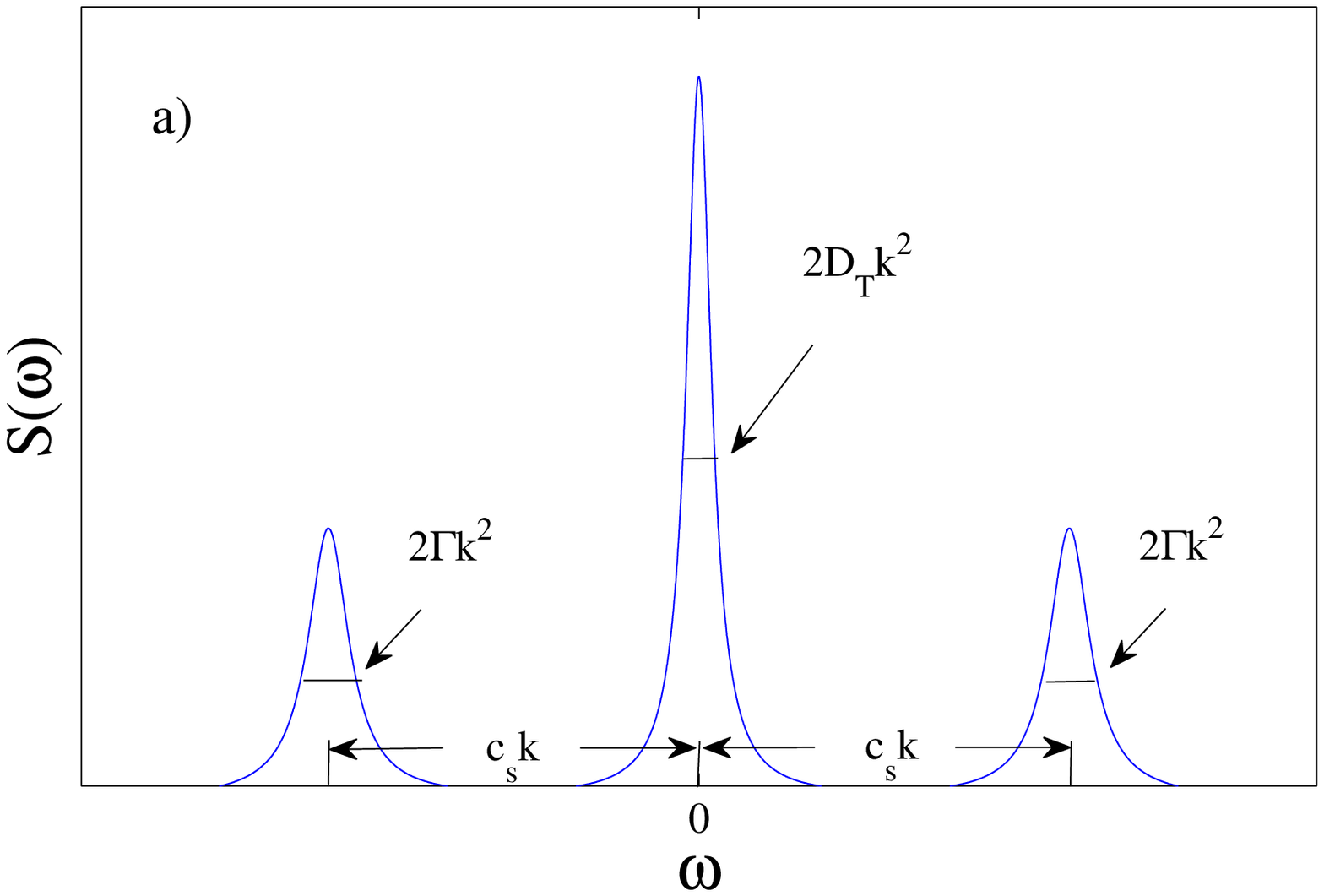,height=8cm,angle=0}} \vskip
		-0.3cm \centerline{
			\epsfig{figure=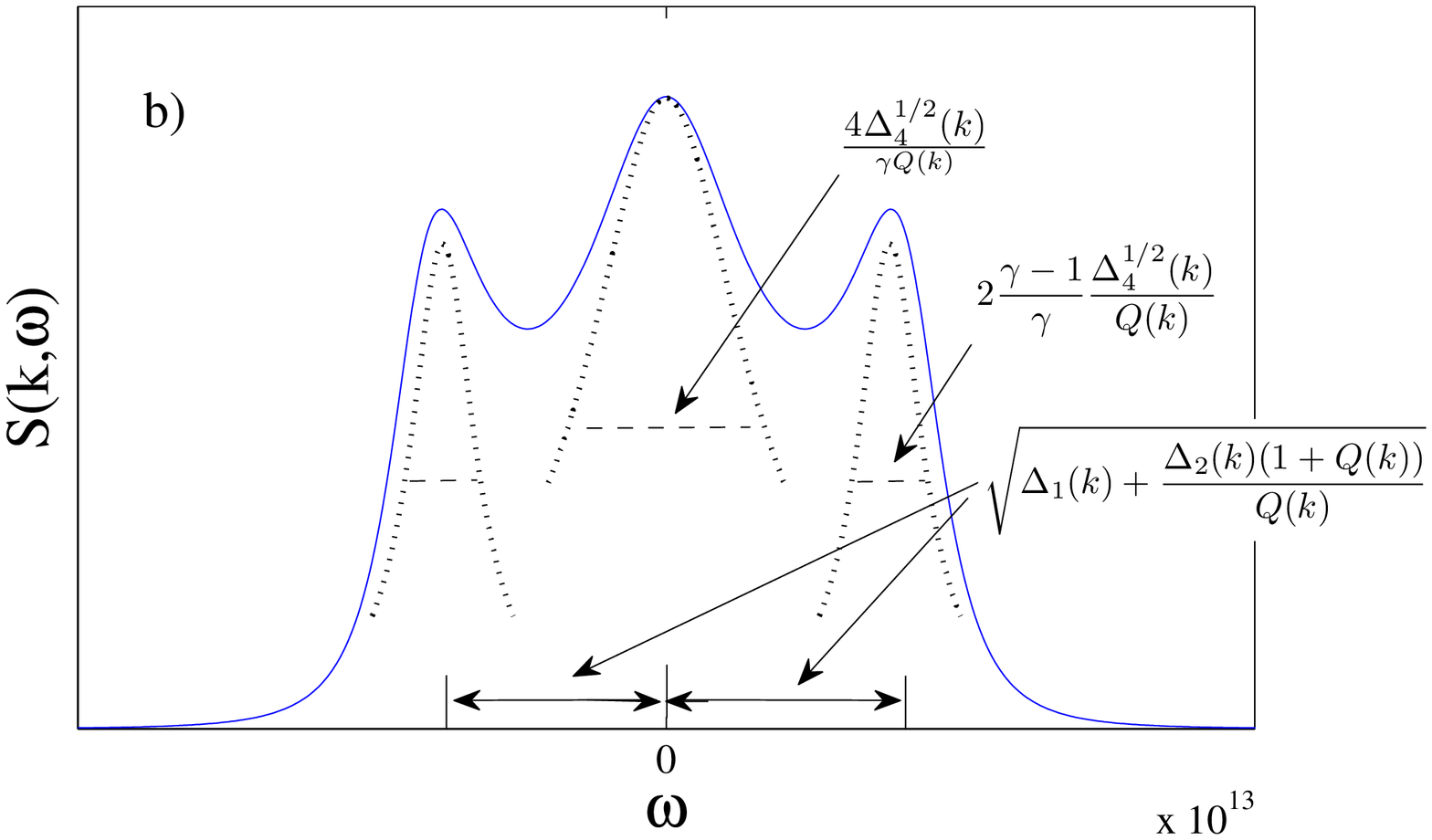,height=8.5cm,angle=0}}
		\caption{(a)  A typical spectrum of the dynamical structure factor recorded in experiments on the
			scattering of light in a simple liquid~\cite{Lansberg_optics,Fabel} --  which corresponds to the hydrodynamic domain with
			small values of $k$; (b)  A dynamical structure factor spectrum recorded in experiments on inelastic
			scattering of slow neutrons and X-rays~\cite{Novikov_2011_1,Novikov_2011_2}  which corresponds to microscopic spatial scales with
			finite values of $k$. The positions and widths of the side peaks are determined by expressions (\ref{eq: side peak
				pos}) and (\ref{eq: side peak}) The half-width of the central peak is given by (\ref{eq: central peak}). }
		\label{fig1}
	\end{figure}

	1. The domain of the transition to the hydrodynamic limit satisfies the condition
	\[
	|\xi(k)| \ll 1,
	\]
	 which
	allows including the domain of conventionally small frequencies (large time scales). The dispersion
	equation then becomes
	\begin{eqnarray}
	\label{cub} s^3+ \frac{2\Delta_4^{1/2}(k)}{\mathcal{Q}(k)}s^2
	&+&\left [\Delta_1(k)+
	\frac{\Delta_2(k)(1+\mathcal{Q}(k))}{\mathcal{Q}(k)}\right ]s +
	\nonumber\\ &+&
	\frac{2\Delta_4^{1/2}(k)\Delta_1(k)}{\mathcal{Q}(k)}=0.
	\end{eqnarray}
	Solving this equation in accordance with the Mountain scheme ~\cite{Mountain_RMP_1966},  we obtain approximate solutions
	of the form
	\begin{eqnarray} \label{eq: disp_om}
	s_{1,2}(k)&=&\pm ic_sk
	-\Gamma k^2,\\
	s_{3}(k)&=&- D_T k^2,\nonumber
	\end{eqnarray}
	where the adiabatic speed of sound $c_s$,  the sound attenuation coefficient $\Gamma$, the temperature conductance $D_T$ and the ratio of specific heat capacities $\gamma=C_p/C_v$ are related by the expressions
	\begin{equation}
	c_s=\sqrt{\gamma} c_{0}, \ \ \ \  \Delta_1(k) = c_{0}^2  k^2,
	\label{hydr_om1}
	\end{equation}
	\begin{equation}
	1+\frac{\Delta_2(k)[1+\mathcal{Q}(k)]}
	{\Delta_1(k)\mathcal{Q}(k)} 
	= \gamma,
	\end{equation}
	\begin{equation} \label{eq: side peak pos}
	\Delta_1(k) + \frac{\Delta_2(k)(1+\mathcal{Q}(k))}{\mathcal{Q}(k)}
	=
	c_s^2 k^2,
	\end{equation}
	\begin{equation} \label{eq: side peak}
	\frac{\gamma-1}{\gamma}
	\frac{\Delta_4^{1/2}(k)}{\mathcal{Q}(k)} 
	= \Gamma k^2,
	\end{equation}
	\begin{equation} \label{eq: central peak}
	2\frac{\Delta_4^{1/2}(k)}{\gamma
		\mathcal{Q}(k)}
	= D_T k^2,
	\end{equation}
where $c_0$ is the isothermal speed of sound.
	Expressions~(\ref{hydr_om1})in the limit of small wave numbers $k$ correspond to the results of the hydrodynamic Landau–Placzek theory~\cite{Landau}.  The real and imaginary
	parts of the first two solutions in \Ref{eq: disp_om}  determine the position and width of the Mandelshtam–Brillouin
	doublet (see Fig.~\ref{fig1}).\footnote{We note that approximate solutions~(\ref{hydr_om1}) are applicable when the ratio between $\Delta_4(k)$ and $\Delta_3(k)$ is large compared
		with the ratios between  $\Delta_1(k)$, $\Delta_2(k)$ and
		$\Delta_3(k)$. This condition is related to the divergence of frequency parameters in the
		hydrodynamic limit, which allows passing to the Mountain procedure~\cite{Mountain_RMP_1966}.}
	\vskip 0.5cm
	
	2.  We consider the high-frequency domain (independent of the wave number $k$ values) defined by the
	condition $s^2/\Delta_4(k)\gg
	1$. In this case, the dispersion equation has the solutions
	\begin{eqnarray}
	s_{1,2}(k) \approx \pm i \sqrt{\Delta_1(k)+ \Delta_2(k)} \equiv
	\pm i \omega_L(k), \label{solid}
	\end{eqnarray}
which reproduce the usual "instantaneous" solid-body response~\cite{Balucani_Zoppi}.
	
	As we can see from dispersion equation~(\ref{condd}) and the obtained solutions (\ref{hydr_om1})--(\ref{eq: central peak}),  the width and
	position of side peaks in the spectra are completely determined by the four frequency parameters $\Delta_j(k)$ ($j=1,\;2,\;3$ and $4$). This is a direct indication that high-frequency spectral properties on microscopic spatial scales
	are determined by two-, three-, and four-particle correlations. Although the high-frequency dynamics can be
	considered a manifestation of solid-body properties, sufficiently large lifetimes of high-frequency excitations
	(compared with solid-body ones) are a characteristic feature of the dynamics of a liquid, where two-,
	three-, and four-particle correlations are pronounced on spatial scales comparable to atomic and molecular
	sizes~\cite{Mokshin_JPCM_2006}.
	
	\subsection{Scaling in relaxation. \label{sec: scaling}}
	
	In the case where the relaxation of various processes exhibits scaling
	expressed by a relation of form~(\ref{eq:
		scaling}), the corresponding theoretical description can be realized in the framework of the self-consistent approach presented here. Theoretical models developed in the mode-coupling
	approximation do belong to this class.~\ref{sec: selfconsist}.
For example, if a scaling relation is satisfied in the relaxation processes associated with the variables
 $A_{\nu}$ and
	$A_{\nu+1}$ of the set $\mathbf{A}$,
	\begin{eqnarray}
	\phi_{\nu+1}(t)&=& \sum_i  \mathcal{A}_i \phi_{\nu}(t)^{p_i},\\
	\sum_i \mathcal{A}_i &=&1, \ \ \ \nu \geq 0,\;p_i>0, \nonumber
	\end{eqnarray}
then the $\nu$-th equation in chain ~(\ref{eq: chain_TCF}) becomes:
	\begin{equation} \label{eq: numer_sol}
	\frac{d\phi_{\nu}(t)}{dt} = \Delta_{\nu+1}\int_{0}^{\infty}
	\phi_{\nu}(t-\tau) \left [ \sum_i  \mathcal{A}_i
	\phi_{\nu}(t)^{p_i} \right ] d\tau .
	\end{equation}
	An equation of form~(\ref{eq: numer_sol}) can be solved numerically with the initial condition [cf. (\ref{prop_TCF})-(\ref{eq: deriv})]:
	\begin{equation} \phi_{\nu}(t=0)=1,\ \ \
	\left. \frac{d\phi_{\nu}(t)}{dt}\right |_{t=0}=0.
	\end{equation}
	
	\vskip 0.5cm \textbf{1. Structure relaxation in supercooled liquids: Mode-coupling approximation.}
	
We
choose the original dynamical variable as
	\begin{equation}
	A_0(t) \equiv \rho_{\vec{k}}^{(s)}(t) = \exp[i \vec{k}\cdot
	\vec{r}_s(t)],
	\end{equation}
	which is the Fourier component of $\rho_s(\vec{r},t)
	= \delta[\vec{r}-\vec{r}_s(t)]$  (see expression~(\ref{eq: traj})) and whose TCF is the so-called
	scattering function~\cite{Hansen/McDonald_book_2006}
	\begin{equation}
	\phi_0(k,t) \equiv F_s(k,t) = \langle \exp\left [ i \vec{k}
	(\vec{r}_s(t) - \vec{r}_s(0))  \right ] \rangle .
	\end{equation}
	Here,
	\begin{equation}
	S_s(k,\omega) = \frac{1}{2\pi} \int_{-\infty}^{\infty}
	\mathrm{e}^{-i\omega t}F_s(k,t) dt
	\end{equation}
is the noncoherent component of inelastic scattering of neutrons and contains information about the oneparticle dynamics.
	\begin{figure}[h!] \vskip -0.5cm
		\centerline{\epsfig{figure=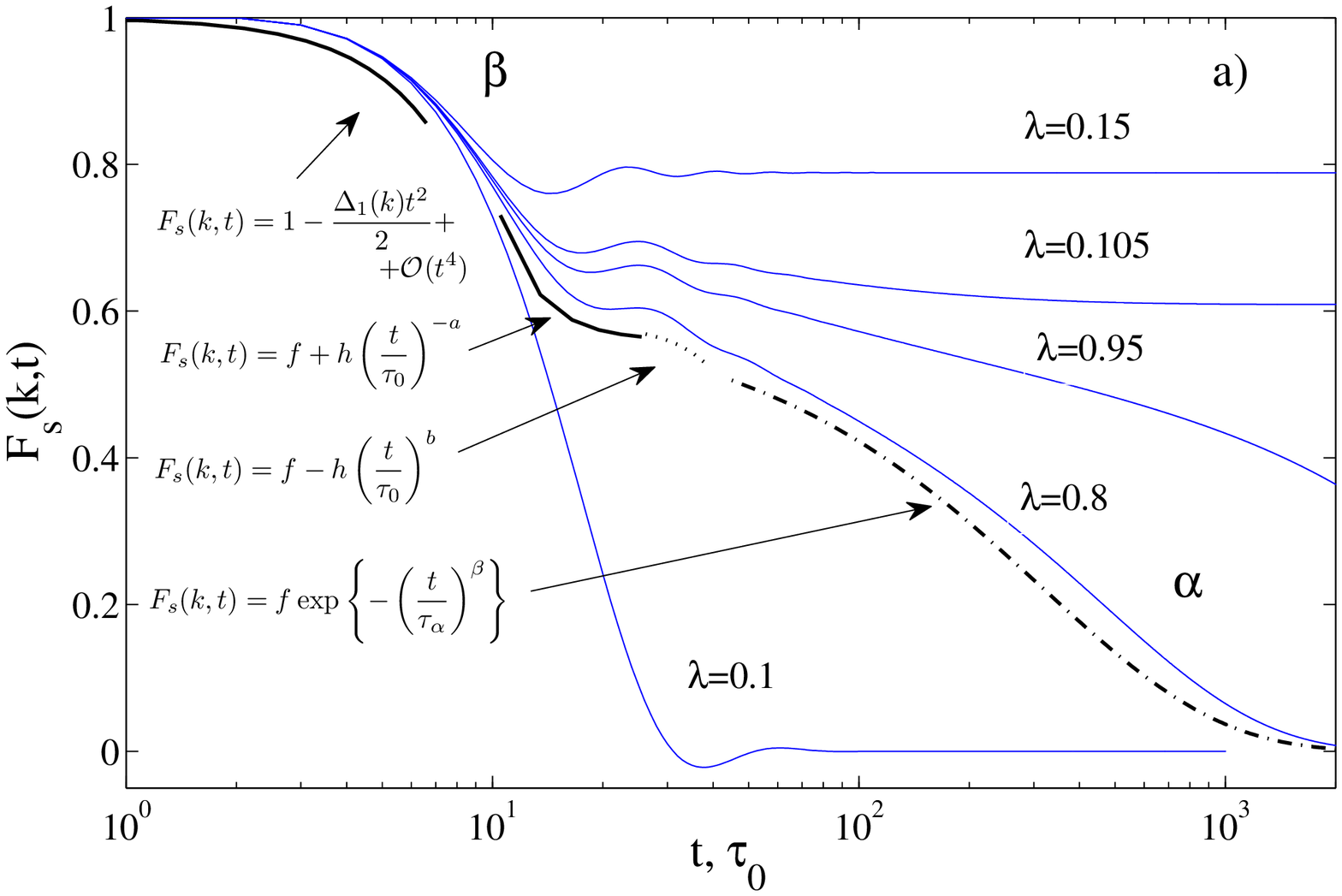,height=9cm,angle=0}}
		\centerline{\epsfig{figure=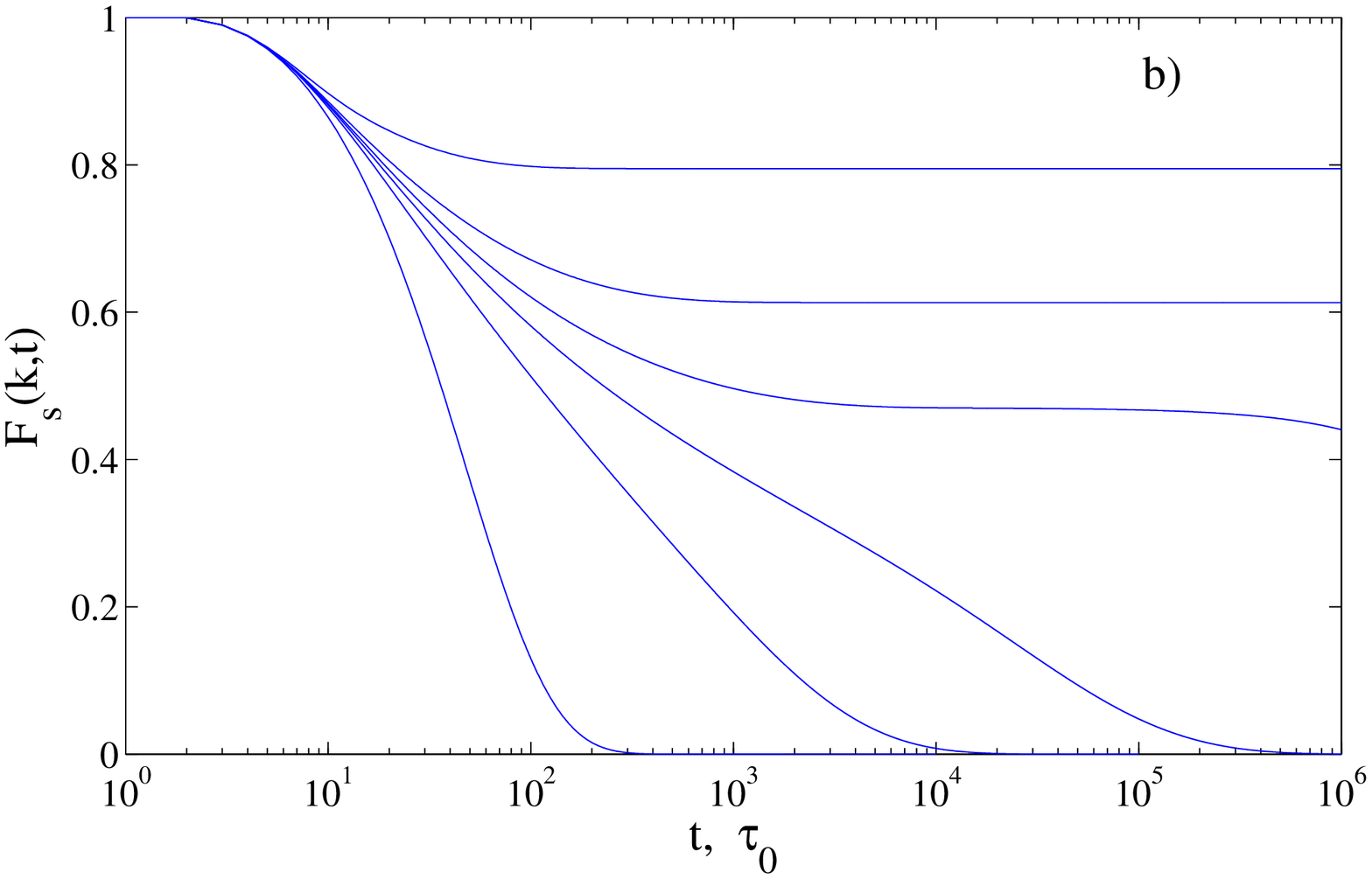,height=9cm,angle=0}}
		\caption{ The scattering function $F_s(k,t)$ obtained by solving Eq.~(\ref{GLE_TCF}) numerically in two different
			approximations of form~(\ref{eq: MCT_gen}): the results in the case of (a) approximation(\ref{eq: Leuth}) and b) approximation~(\ref{eq: Gotze}). Here $\lambda =
			\Delta_2(k)/4\Delta_1(k)$ and $\tau_0$  is the unit time scale.} \label{fig2}
	\end{figure}
	
	We write the first two equations of chain~(\ref{eq: chain_TCF}):
	\begin{equation} \label{eq: 1}
	\frac{dF_s(k,t)}{dt} = - \Delta_{1}(k) \int_0^{\infty}
	F_s(k,t-\tau) \phi_{1}^{(s)}(k,\tau) d\tau ,
	\end{equation}
	\begin{equation} \label{eq: 2}
	\frac{d\phi_{1}(k,t)}{dt} = - \Delta_{2}(k) \int_0^{\infty}
	\phi_{1}^{(s)}(k,t-\tau) \phi_{2}^{(s)}(k,\tau) d\tau ,
	\end{equation}
	where the frequency parameter $\Delta_1(k) = k_B T k^2/m$, now characterizes the thermal motion of the particle,
	\begin{equation}
	\Delta_2(k) = 2\Delta_1(k) + \frac{\langle \nabla^2 U(\vec{r}_s)
		\rangle}{3m} + \mathcal{O}(k^2),\ \ \  |\vec{r}_s| \simeq 2\pi/k,
	\end{equation}
	the first-order memory function $\phi_1^{(s)}(k,t)$  is identified with the TCF of the longitudinal velocity component
	of the particle, and the memory function
	$\phi_2^{(s)}(k,t)$ is identified with the TCF of the force $\vec{f}_s$, acting on the
	particle from the environment. Combining (\ref{eq: 1}) and (\ref{eq:
		2}), we obtain a generalized Langevin equation in
	the form ~\cite{AVM_CP_2007}
	\begin{equation} \label{GLE_TCF}
	\frac{d^2 F_s(k,t)}{dt^2} + \Delta_1(k) F_s(k,t) + \Delta_2(k)
	\int_0^{t} d\tau \; \phi_2^{(s)}(k,t - \tau) \frac{d
		F_s(k,\tau)}{d \tau} = 0,
	\end{equation}
	solving which requires knowing the behavior of the memory function
	$\phi_2^{(s)}(k,t)$. Following the key idea of
	the mode-coupling approximation on the correlation between processes ~\cite{Kawasaki_1},  associated with the stochastic
	force
	$f_s$ and those associated with the structure relaxation directly, we can write the scaling expression
	\begin{eqnarray} \label{eq: MCT_gen}
	\phi_2^{(s)}(k,t) &=& \mathcal{A}_1 F_s(k,t) + \mathcal{A}_2
	F_s(k,t)^p, \\   0 &\leq& \mathcal{A}_1,\mathcal{A}_2 \leq 1,\ \ \
	\mathcal{A}_1+\mathcal{A}_2=1,\ \ \ p>1, \nonumber
	\end{eqnarray}
	where $\mathcal{A}_1$ and $\mathcal{A}_2$ are weight coefficients. Substituting expression~(\ref{eq: MCT_gen}) in (\ref{GLE_TCF}),
we obtain an equation
that can be solved numerically and gives a variety of solutions suitable for reproducing the properties of
the noncoherent scattering function $F_s(k,t)$ in the "supercooled liquid–glass" transition domain. As an
example, we show solutions for $F_s(k,t)$, obtained from~(\ref{GLE_TCF}) in two particular cases in~\ref{fig2}
	
	1. expression~(\ref{eq: MCT_gen}) has the form
	\begin{equation} \label{eq: Leuth}
	\phi_2^{(s)}(k,t) = F_s(k,t)^2,
	\end{equation}
	which is a simplified Leutheusser model
	~\cite{Leutheusser_1984,Goetze} [see
	~\ref{fig1}a)] and

	2.  expression~(\ref{eq: MCT_gen}) can be represented as
	\begin{equation} \label{eq: Gotze}
	\phi_2^{(s)}(k,t) = \mathcal{A}_1 F_s(k,t) + \mathcal{A}_2
	F_s(k,t)^2,
	\end{equation}
	which is a simplified G\"{o}tze–Sjogren  $M_{12}$-model
	~\cite{Gotze_2009} [see~\ref{fig2}b)].
	
	As can be seen from~\ref{fig2}, both models reproduce a transition of the system from the ergodic to
	nonergodic phase, the transition associated with the appearance of a plateau in the time dependence $F_s(k,t)$.
Under this transition, the scattering function starts being characterized by two-stage relaxation,
under which fast relaxation processes mainly associated with the vibrational dynamics of particles are
designated as $\beta$ relaxation and slow processes responsible for the structure transformations are associated
with
	$\alpha$ relaxation~\cite{Hansen/McDonald_book_2006}.  The transition to the nonergodic phase is here governed by the frequency parameters
	$\Delta_1(k)$ and $\Delta_2(k)$.  In the example of model~(\ref{eq:
		Leuth}) the transition can be conveniently characterized by
	introducing the parameter
	$\lambda = \Delta_2(k)/4\Delta_1(k)$. The occurrence of the transition is then associated with the
	condition $\lambda=1$ [see Fig.~\ref{fig2}a)].
	
An important property of the mode-coupling approximation is that it allows reproducing characteristic
stages in the temporal behavior of the TCF for a supercooled liquid. For example, the behavior of
	$F_s(k,t)$ at the initial stage of $\beta$ relaxation is characterized by the dependence
	\begin{equation} \label{eq: short_time}
	F_s(k,t) \simeq 1 - \frac{\Delta_1(k)}{2}t^2 + \mathcal{O}(t^4),
	\end{equation}
	which follows from~(\ref{taylor}).  Next, the final stage in
	$\beta$ relaxation, preceding the formation of a plateau in $F_s(k,t)$
	[see Fig.~\ref{fig2}1)], is described by the so-called critical law
	\begin{equation}
	F_s(k,t)\simeq f^{(s)}(k)+h(k)\left ( \frac{t}{\tau_0(k)}\right
	)^{-a},
	\end{equation}
	where $f^{(s)}(k)$  is the nonergodicity parameter characterizing the height of the plateau in $F_s(k,t)$ and $\tau_0(k)$  is
	the relaxation scale. In accordance with approximation ~(\ref{eq: Leuth})
 the exponent in the critical law is determined
from the expression
	\begin{equation}
	\frac{\Gamma^2(1-a)}{\Gamma(1-2a)} =
	\frac{\Delta_2(k)}{\Delta_1(k)}[1-f^{(s)}(k)]^3,
	\end{equation}
	where $\Gamma(\ldots)$ is the gamma function. The “initial” stage of $\alpha$ relaxation [see Fig.~\ref{fig2}b)],  characterizing the
	subsequent attenuation of the TCF after the steady behavior with
	$F_s(k,t)\simeq f^{(s)}(k)$, occurs in accordance
	with the von Schweidler relaxation~\cite{Gotze_2009}:
	\begin{equation}
	F_s(k,t) \simeq f^{(s)}(k) - h(k)\left ( \frac{t}{\tau_0(k)}\right
	)^{b},
	\end{equation}
where in the case of approximation~(\ref{eq: Leuth}) the exponent $b$ is related to the frequency parameters as
	\begin{equation}
	\frac{\Gamma^2(b+1)}{\Gamma(2b+1)} =
	\frac{\Delta_2(k)}{\Delta_1(k)}[1-f^{(s)}(k)]^3.
	\end{equation}

	Finally, we note that at large time scales corresponding to structure relaxation scales, the temporal
	behavior of the TCF $F_s(k,t)$  obtained from~(\ref{eq:
		MCT_gen}) allows approximately reproducing the relaxation behavior
	described by the stretched exponential (Kohlrausch function),
	\begin{equation}
	F_s(k,t) \simeq f^{(s)}(k)\exp\left [ - \left (
	\frac{t}{\tau_{\alpha}} \right )^{\beta} \right ],
	\end{equation}
	where $\beta < 1$ and $\tau_{\alpha}$ is identified with the structure relaxation time.
	
	\vskip 1cm
	
\textbf{Acknowledgments.} The author thanks Kamilla Ishdavletova (KFU, Kazan, Russia) for help in typing the text of this work. The author is grateful to Ramil Khusnutdinoff (KFU, Kazan, Russia) for the useful
comments and help in numerical computations leading to the results presented in Fig.~\ref{fig2} and to Arkady Novikov (Leipunsky PEI, Obninsk, Russia), Peter H\"{a}nggi (Universit\"{a}t Augsburg, Augsburg, Germany), Jean-Louis Barrat (Universit\'{e} Joseph 	Fourier, Grenoble, France), and M. Howard Lee (University of Georgia, Athens, USA) for the numerous discussions of the problems touched upon in this paper. The author sincerely thanks Valentin Ryzhov and Vadim Brazhkin for the useful questions, suggestions, and discussions of the results in this paper at the seminars in the IHPP, Russian Academy of Sciences, in 2013–2014.
	
This research was supported by the Russian Science Foundation (Grant No. 14-12-01185).

\end{document}